\documentclass[aps,prL,twocolumn,superscriptaddress,showpacs]{revtex4-1}
\usepackage{graphicx}
\usepackage[dvipsnames,usenames]{color}
\usepackage{amsmath} 
\usepackage{amssymb}
\usepackage{bm}
\usepackage{multirow}
\usepackage{wasysym}
\usepackage{subfigure}
\usepackage{verbatim}
\usepackage{marvosym}
\usepackage{latexsym}
\usepackage{xcolor}
\usepackage[dvipsnames,usenames]{color}
\usepackage{float} 

\usepackage{xr}
\makeatletter
\newcommand*{\addFileDependency}[1]{
  \typeout{(#1)}
  \@addtofilelist{#1}
  \IfFileExists{#1}{}{\typeout{No file #1.}}
}
\makeatother

\newcommand*{\myexternaldocument}[1]{
    \externaldocument{#1}
    \addFileDependency{#1.tex}
    \addFileDependency{#1.aux}
}

\myexternaldocument{SI}

\begin{document}
\title{Soft ULC Microgels at the Interface Interact and Flow as Hertzian-like Colloids}

\author{Jos\'e Ruiz-Franco}
\email{These authors contributed equally to this work.}
\affiliation{Department of Condensed Matter Physics, University of Barcelona, 08028 Barcelona, Spain}
\affiliation{Institute of Complex Systems (UBICS), University of Barcelona, 08028 Barcelona, Spain}

\author{Tom H\"ofken~$^*$}
\affiliation{Institute of Physical Chemistry, RWTH Aachen University, Landoltweg 2, 52074 Aachen, Germany}

\author{Maximilian M. Schmidt}
\affiliation{Institute of Physical Chemistry, RWTH Aachen University, Landoltweg 2, 52074 Aachen, Germany}

\author{Steffen Bochenek}
\affiliation{Institute of Physical Chemistry, RWTH Aachen University, Landoltweg 2, 52074 Aachen, Germany}

\author{Emanuela Zaccarelli}
\email{emanuela.zaccarelli@cnr.it}
\affiliation{Italian National Research Council - Institute for Complex Systems (CNR-ISC), Sapienza University of Rome, Piazzale Aldo Moro 5, 00185 Rome, Italy}
\affiliation{Department of Physics, Sapienza University of Rome, Piazzale Aldo Moro 2, 00185 Rome, Italy}

\author{Andrea Scotti}
\email{andrea.scotti@fkem1.lu.se}
\affiliation{Division of Physical Chemistry, Lund University, SE-22100 Lund, Sweden}

\date{\today}
\begin{abstract}
Soft pair potentials predict a reentrant liquid phase for high concentrations, a behavior not observed experimentally.
Here, very soft microgels confined at an oil-water interface are used as a model system of particles interacting \emph{via} a soft potential in two dimensions (2D).
Interfacial rheology measurements demonstrate the existence of different flow regimes that depend on the compression of the monolayer.
Such a compression also leads to a non-monotonic variation of the elastic moduli and of the yield stress of the monolayer.
These results, together with the equilibrium phase behavior of the monolayer, are reproduced in molecular dynamics simulations of a 2D system of particles interacting with a Hertzian-like potential.
Remarkably, due to the non-monotonic variation of the elastic moduli, we observe \textit{isoelastic} points where the monolayer shows the same stiffness at very different concentrations.
These points are the experimental manifestation of the predicted reentrant liquid phase. 

\end{abstract}

\maketitle 

\section{Introduction}
Softness can be defined as the tendency of a system to respond to small changes of external parameters with great variations of a physical quantity \cite{deG05}.
A textbook example is represented by the viscosity of complex fluids, which increases by several orders of magnitude when the concentration of the building blocks suspended in the solvent overcomes a threshold value \cite{Seg95}.
For hard spheres, this critical concentration corresponds to a volume fraction of $\phi \sim 0.58$, where the glass transition takes place \cite{van1994glass}.
When the building blocks are soft particles, they can deform and decrease in volume as their concentration increases, which changes the viscosity and, more generally, the flow properties of the suspension both in bulk \cite{Vla14, Sco22_review} and at the interface \cite{Sch23}.
This is the case, for instance, in star polymers \cite{Erw10} or microgels \cite{Bor99, Sco20_flow}, that are polymeric spheres in the colloidal domain. 

When $N$-isopropylacrylamide (NIPAM) is used as the monomer, the microgel polymeric network can form even without the addition of any crosslinker agent during the precipitation polymerisation synthesis \cite{Gao03}.
This happens because of a hydrogen abstraction reaction at the tertiary carbon of the isopropyl group \cite{Bru19}, which leads to rare self-crosslinking events between the NIPAM monomers.
These ultra-low crosslinked (ULC) microgels are the softest pNIPAM-based microgels that can be obtained by precipitation polymerisation \cite{Hou22, Hof24}.
Concentrated suspensions of ULC microgels show a very gradual increase in suspension viscosity with concentration, resembling the behavior of macromolecules and linear polymers rather than that of colloidal particles \cite{Sco20_flow}.
However, these microgels also crystallize similarly to regular microgels \cite{Sco21, Sco22_review} and hard spheres \cite{Pus86}.
The combination of these peculiar features makes them a fascinating system at the border between polymers and colloids \cite{Sco19}.

Their behavior at liquid-liquid interfaces is also very interesting. 
Indeed, once ULC microgels are confined to an interface, they experience a significant lateral deformation assuming a pancake-like structure \cite{Sch21}: they adopt a homogeneous, flat structure unlike microgels with added crosslinker, which retain a sizable core at interfaces, usually referred as a ``fried-egg'' shape. 
Notably, this unusual conformation influences the interaction between interfaces themselves, as ULC microgels can facilitate the creation of smart dispersed emulsions that are susceptible to disruption by external stimuli, in contrast to conventional microgels, which lack this capability. \cite{Rey23}
As a consequence of their high deformability, ULC microgels show, even at interfaces, properties that lie between those of macromolecules and colloidal particles \cite{Sco19, Boc22, Pet23} and remain strongly anchored to the interface upon lateral compression \cite{Ger24}.

To properly understand and predict the properties of these microgels at the interface, one should describe their interaction potential in two dimensions (2D).
Starting with standard microgels, it has previously been shown that a Hertzian potential is able to describe their behavior in bulk only for low enough generalized volume fractions~\cite{paloli2013fluid,del2024numerical}.
However, a multi-Hertzian model is found to improve the description even in the presence of depletion interactions \cite{bergman2018new}.
Moving to standard microgels at interfaces, again the Hertzian model is found to work at low enough generalized area fractions, but in order to interpret interfacial rheological measurements, we proposed a ``square-shoulder-Hertzian'' (SSH) potential \cite{Sch23}.
This model consists of a Hertzian contribution to describe the corona-corona interactions, coupled with a square shoulder potential accounting for the presence of an almost incompressible core in the microgel center \cite{Hou22, Sch21}.
Importantly, this inner core, amounting to roughly 25\% of the total size of the microgel, was found to be much smaller than the core that would be detected in bulk analyzing the microgel form factor measured with small-angle scattering and using the fuzzy sphere model \cite{Hou22, Sti04}. 
This can be either due to the stretching of the microgel at the interface, or because in very concentrated conditions the outer chains sum up to the inner ones, finally making the microgel incompressible.

Turning to ULC microgels, in bulk they can be compressed up to 80\% of their original volume \cite{Hou22}, due to their very low and uniform internal polymer concentration \cite{Sco19, Sch21}.
Thus, the incompressibility detected for harder microgels is expected to be lost and, once at the interface, they would behave as soft homogeneous discs that can be compressed almost indefinitely.
In other words, they can be considered a good model system for soft particles interacting with a purely Hertzian potential.
Until now, experimental systems with such potentials were only observed at low concentrations, while highly concentrated systems of Hertzian-like discs remained purely theoretical.
Consequently, the present investigation of dense suspensions of ULC microgels at the interface fills a gap in the existing knowledge and might also serve to demonstrate the occurrence of an elusive property of soft pair potentials: the existence of a reentrant liquid phase at high concentrations that has never been observed experimentally.

In the present work, we thus extensively investigate the experimental phase behavior and flow properties of dense monolayers of ULC microgels, that are probed by a combination of \emph{ex situ} atomic force microscopy (AFM) of the deposited dry monolayer on solid substrate and \emph{in situ} interfacial rheology of the microgels at the oil-water interface.
The experimental results in equilibrium and under flow are then compared with computer simulations in equilibrium and under shear in which the particles interact with a Hertzian style potential, demonstrating that ULC microgels at interfaces can be accurately described using a purely Hertzian interaction, without the need to introduce a square-shoulder term.
This not only validates the theoretical predictions but also establishes a minimal yet comprehensive model for ultra-soft particle interactions. Indeed, contrary to the case of hard microgels, a square-shoulder term is not needed to reproduce the evolution of the radial distribution function with increasing microgel concentration.

Our findings show that, while a simple Hertzian potential suffices to describe the static features of the monolayer in equilibrium, capturing the dynamical response under shear requires the use of a multi-Hertzian model~\cite{bergman2018new}. 
This highlights the critical role of internal elasticity gradients even in ultra-soft microgels, a factor often overlooked in previous studies.
This amounts to complement the corona-corona interactions with additional Hertzian contributions, representing core-corona and core-core interactions, respectively, since the internal elasticity of standard microgels becomes progressively larger towards the center of the microgel due to the accumulation of polymeric chains \cite{Hou22, Hof24}.
For ULC microgels, having a much more homogeneous density profile~\cite{Haz23} due to the absence of crosslinkers, they still experience an increasing effective elasticity as a function of compression \cite{Hou22, Hof24}.
Hence, in this case, the use of a multi-Hertzian model complements the interactions between two isolated ULC microgels adsorbed at the interface \cite{Cam20} by incorporating the effect of the surrounding monolayer.
Incorporating these contributions results in a simple pair potential form that effectively encompasses both static structure and mechanical properties under flow conditions for all concentrations.

\begin{figure*}[htbp!]
\centering
  \includegraphics[width=\linewidth]{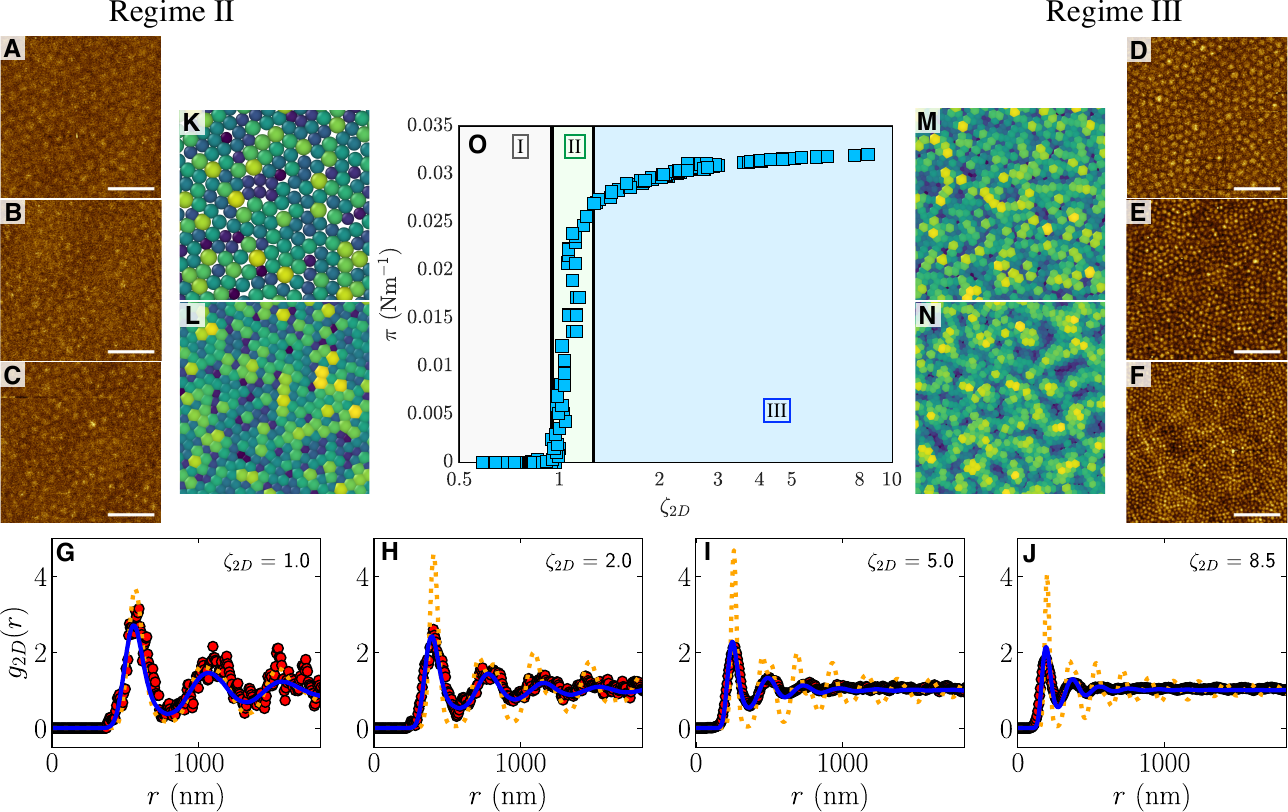}
  \caption{\textbf{Equilibrium phase behavior of ULC microgels at the interface.} (\textbf{A-F}) Atomic force microscopy micrographs of the monolayer of ULC microgels deposited on a solid substrate. The generalized area fraction $\zeta_{2D}$ values equal (\textbf{A}) $0.99\pm0.21$, (\textbf{B}) $1.07\pm0.22$, (\textbf{C}) $1.43\pm0.30$, (\textbf{D}) $2.05\pm0.42$, (\textbf{E}) $5.11\pm1.05$ and (\textbf{F}) $8.48\pm1.74$. (\textbf{G-J}) Radial distribution functions $g(r)$ computed from the AFM micrographs (symbols) and from simulations using a Hertzian (solid line) or a multi-Hertzian (dotted line) pair potential. (\textbf{K-N}) Simulation snapshots for the shown $g(r)$, where differently colored particles indicate the size-polydispersity (from blue to yellow increasing particle size). (\textbf{O}) Surface pressure, $\pi$, as a function of $\zeta_{2D}$ highlighting the three different regions (I-III) examined throughout the text in different colors.}
  \label{fig:ULC3Gr}
\end{figure*}

Experimentally, the elastic modulus and apparent yield stress varies as a function of compression with a non-monotonic behavior.
This property and the glassy nature of the interface are accurately reproduced by the use of the multi-Hertzian model.
Such a non-monotonic variation and the presence of \emph{isoelastic points}, where the monolayer present the same stiffness at very different concentrations, are interpreted as the experimental manifestation of the reentrant liquid phase predicted theoretically for such soft pair potentials \cite{Cam20, Ber10}.
Furthermore, from the experimental results, we identify two main flow regimes, characterized by two distinct master curves and providing a unified framework that extends the understanding of microgels at interfaces beyond the well-established behavior of hard microgels \cite{Sch23}. 
This suggests that such a behavior is generic for microgels at interfaces, independently of the crosslinker concentration and might pave the way to a general description of the phase behavior and flow properties of other soft colloids at the interface. 
This might be of interest for complex biological matter, such as globular proteins or antibodies, since these are found to aggregate at surfaces over time.  \cite{Woo23,Fre04}
In the literature, it has been observed that the adsorption of proteins at the interface can significantly alter the viscoelatic properties of the system. 
For instance, the corse of the surface pressures, and of the elastic modulus of interfaces with adsorbed proteins show non-monotonic behavior that varies significantly depending on the specific protein involved. \cite{Bos01} 
For such systems, a complex interplay between soft repulsion and electrostatic interactions dominate the interfacial response.
Using microgel monolayers as discussed here provides a simplified model system to disentangle the specific contribution of softness to the evolution of interfacial elasticity with increasing concentration, while also providing a clear physical picture that can be later incorporated into the understanding of more complex models involving additional interactions.

\section{Results}
\subsection{From structure and dynamics to pair interactions}

First, we examine the equilibrium phase behavior of ULC microgels as a function of generalized area fraction, $\zeta_{2D}$ (see Eq.~\ref{eq:z2D}) using atomic force microscopy.
Once these microgels adsorb to an interface they deform and resemble very flat disks \cite{Sco19, Boc22,Rey23,Boc23}.
As a consequence of the little resistance that the poorly crosslinked network opposes to this deformation, the size polydispersity is found to significantly increase for the adsorbed microgels with respect to its value in bulk \cite{Sco19}.
This rise in size disparities makes these very soft microgels unable to form crystals at the interface as highlighted by the AFM micrographs and the $g(r)$ reported in Fig.~\ref{fig:ULC3Gr}. 
Despite the AFM micrographs, and the related $g(r)$, are obtained \textit{ex situ}, a recent study  has revealed that for loosely crosslinked microgels there is no difference between \textit{ex situ} and \textit{in situ} structure within the monolayer~\cite{Kuk23}.

\begin{figure}
\includegraphics[width=0.48\textwidth]{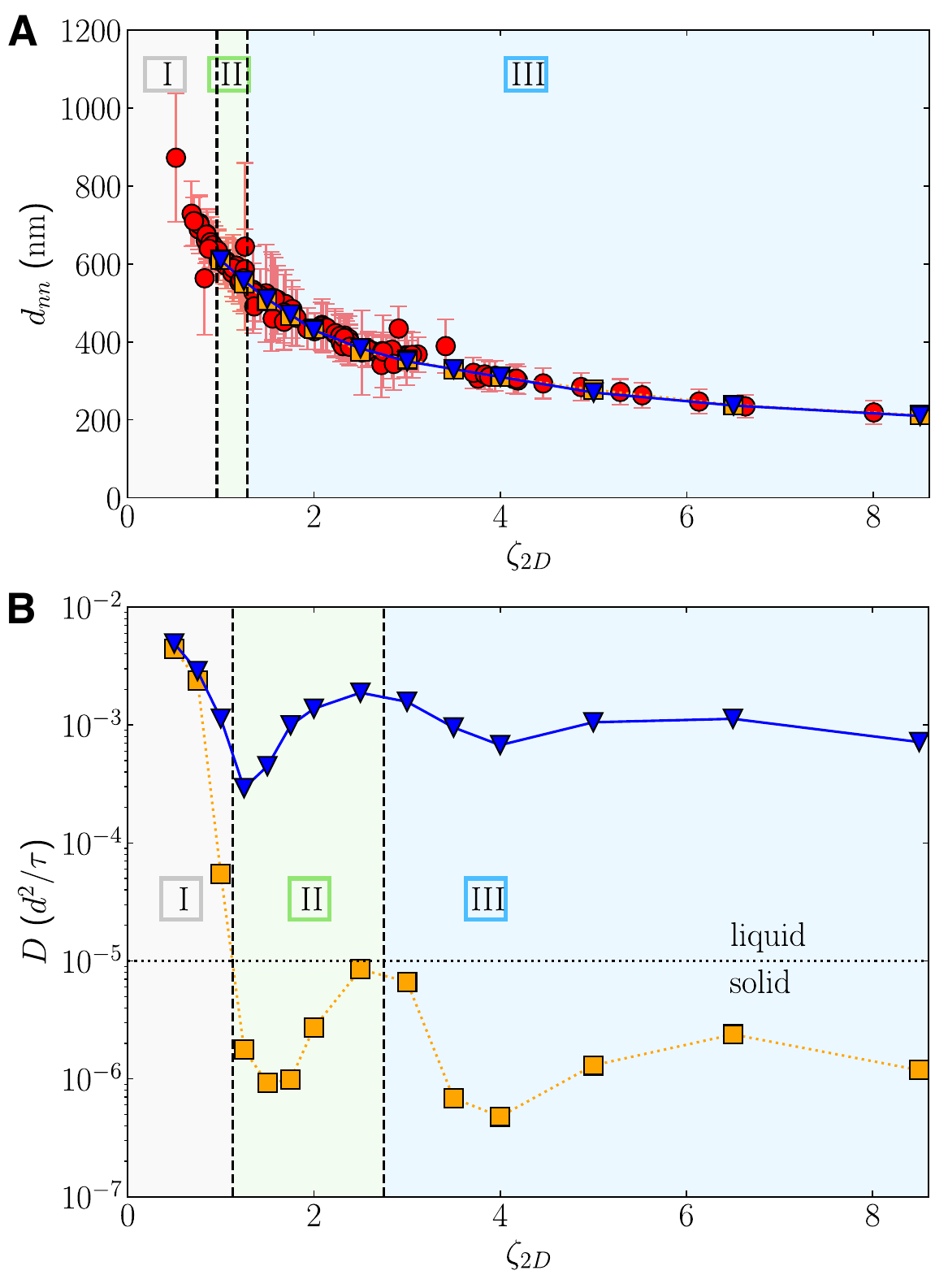 }
\centering
\caption{\textbf{Structure and dynamics properties.} (\textbf{A}) The nearest neighbor distance, $d_{nn}$, as a function of $\zeta_{2D}$. Simulation results for a Hertzian (triangles) and a multi-Hertzian (squares) lay on top of the experimental data (circles). Solid and dotted lines are fits of the data with $d_{nn}\propto\zeta_{2D}^{-1/2}$. (\textbf{B}) Self-diffusion coefficient $D$ calculated from simulations of Hertzian (triangles) and  multi-Hertzian (squares) spheres in equilibrium as a function of $\zeta_{2D}$. Empirically, we find our interfaces with a diffusion coefficient $D>10^{-5}~\sigma^2/\tau$ tend to behave as liquids with zero yield-stress. The different regimes of the compression isotherms in Fig.~\ref{fig:ULC3Gr} are indicated by different colors.}
\label{fig:New2}
\end{figure}

From the analysis of the AFM micrographs in Figs.~\ref{fig:ULC3Gr}A-to-\ref{fig:ULC3Gr}F, we can extract the corresponding $g(r)$, shown as symbols in Figs.~\ref{fig:ULC3Gr}G-to-\ref{fig:ULC3Gr}J. 
For all the measured $\zeta_{2D}$ values (Figs.~\ref{fig:ULC3Gr}A-to-\ref{fig:ULC3Gr}F, the first peak is well-defined but the presence of few, quite smeared additional peaks indicates that the ULC microgels become closer and closer in the monolayer always remaining in a solid disordered phase, (Figs.~\ref{fig:ULC3Gr}I and~\ref{fig:ULC3Gr}J), in contrast to the hexagonal lattice easily formed by hard microgels \cite{Rey16, Sco22_review}.
Furthermore, ULC microgels do not exhibit a significant variation of stiffness within their volume, as previously shown in Ref.~\cite{Sch21}.
This also implies the absence of a second increase in the surface pressure, $\pi$, versus $\zeta_{2D}$ curve, for $\pi>3.0\cdot10^{-2}$~Nm$^{-1}$ \cite{Rey16, Gei14, Boc21}, as shown by the squares in Fig.~\ref{fig:ULC3Gr}O which reveal only three regimes \cite{Sco19}: 
(I) for $\zeta_{2D} \leq 0.95$ a gas-like state where microgels are separated and $\pi = 0$; (II) a region where the polymer progressively reaches the maximum coverage of the surface and microgels are compressed, with $0.95<\zeta_{2D} \leq 1.27$ marking the range where the microgels makes contact and $\pi$ increases; (III) a region where no more polymer can occupy the surface ($\pi\simeq 3.3\cdot10^{-2}$~Nm$^{-1}$ reaches a plateau), and the microgels get progressively more and more compressed as $\zeta_{2D} > 1.27$. 
Despite becoming more compressed and closer together, these  very soft microgels are not significantly altered  while being compressed, i.e.~there seems to be no threshold at which they become significantly stiffer like hard microgels.
A confirmation of this behavior is given by the evolution of the nearest neighbor distance, $d_{nn}$, corresponding to the position of the first peak of the $g(r)$, reported in Fig.~\ref{fig:New2}A as circles, which shows a continuous decrease at all studied $\zeta_{2D}$.
The data closely follow the prediction of a homogeneous filling of the space as $\zeta_{2D}^{-1/2}$ at all investigated generalized area fractions.
This contrasts with observations for hard microgels, where a second, constant length scale was found to emerge at a given $\zeta_{2D}$ \cite{Rey16, Boc21, Sco19, Sco22_review}, denoting the onset of an incompressible core, a feature thus absent in ULC microgels at the interface.

The experimental results are complemented by extensive Brownian dynamics simulations to verify that whether the interactions at the interface between microgel at the equilibrium can be approximated by a Hertzian potential.
In particular, we investigate the response of a model system consisting of 10~\% polydisperse disks with an average diameter $d$ interacting with a soft 2D  Hertzian potential (see Eq.~\ref{eq:2dhertzian}) in equilibrium and under external shear.
This pair potential has already been successfully applied to model the interactions between individual in-silico microgels at the interface \cite{Cam20}.
Representative snapshots of the simulated system are shown in Figs.~\ref{fig:ULC3Gr}K-to-\ref{fig:ULC3Gr}N.
Similarly to what was done for the AFM micrograph, we extract the $g(r)$ corresponding to the simulated system.
We find that the evolution of the radial distribution function in equilibrium aligns with the one observed experimentally across the full range of $\zeta_{2D}$, as demonstrated by comparing the symbols (experimental data) and the solid lines (simulations) in Figs.~\ref{fig:ULC3Gr}G-to-\ref{fig:ULC3Gr}J.
This is further confirmed in Fig.~\ref{fig:New2}A, where the simulated nearest-neighbor distance (triangles) obtained using the 2D Hertzian model is compared against  the experimental data (circles) at all investigated $\zeta_{2D}$.

However, for high density systems, it is well-known from the literature that the Hertzian model predicts a reentrant melting of the solid at low interaction strengths~\cite{Ber10}, a feature also present in 2D~\cite{Cam20}, that so far was not observed experimentally.
Hence, in order to obtain a clear comparison of the rheological properties, it would be more desirable to work with a model system that remains solid for all generalized area fractions above jamming.
Unfortunately, within the limits of the Hertzian potential this cannot be achieved without significantly increasing the stiffness of the discs and, therefore, not representing any longer the behavior of ULC microgels well, particularly mismatching the position of the peak of the $g(r)$.
An improvement is found by using the multi-Hertzian potential (see Eq.~\ref{eq:MH}), which adds up multiple Hertzian terms with different interaction strengths and sizes, as discussed in Methods.
In this way, the properties at low $\zeta_{2D}$ can be maintained by using the same Hertzian potential up to contact and adding shorter-ranged, stiffer Hertzian contributions which influence the behavior of the system at higher compression.
Differently from regularly crosslinked microgels in 3D \cite{bergman2018new}, where the different Hertzian potentials were directly linked to structural elements of the microgel, such as core, shell or corona, the ULC microgels have a more homogeneous internal density distribution and the choice of a multi-Hertzian potential cannot be justified in the same manner.
In this case, the increased stiffness originates from an increasing layer thickness and interpenetration of polymer chains as recently shown by neutron reflectivity experiments \cite{Ger24}, but also it is due to many-body interactions between more than two microgels.
Indeed, these are all phenomena that must be considered in an effective way in a pair potential model.

The multi-Hertzian model is also able to quantitatively reproduce the nearest neighbor distance evolution with $\zeta_{2D}$, as shown by the squares in Fig.~\ref{fig:New2}A.
This implies that the additional stiff terms in the multi-Herztian potential do not influence the spacing between particles, even for high densities.
With increasing $\zeta_{2D}$, the microgels become progressively more compressed, aligning with the trend observed experimentally for the ULC microgels.
However, while the position of the first maximum of the radial distribution function does not change due to the stiffer potential, the height of the peaks increases significantly with $\zeta_{2D}$ as shown by the comparison between the symbols (experimental data) and the dotted lines (simulations) in Figs.~\ref{fig:ULC3Gr}G-to-\ref{fig:ULC3Gr}J. 
We believe that this is the result of introducing the additional stiffness as part of the pair potential and not as a more general many-body term.
Notwithstanding this, the multi-Hertzian model qualitatively captures the experimental $g(r)$ at all investigated generalized area fractions.

From the equilibrium simulations, we can also compute the mean square displacement, $\langle r^2 \rangle$, and extract the diffusion coefficient defined as the change of $\langle r^2 \rangle$ with time after sufficiently long period.
The values of the self-diffusion coefficient $D$, estimated from the long-time limit of the mean-squared displacement, are reported as a function of $\zeta_{2D}$ in Fig.~\ref{fig:New2}B for both Hertzian and multi-Hertzian models, triangles and squares, respectively.
Empirically, we estimate that this particular system becomes glassy when $D < 10^{-5}~\sigma^2/\tau$ (dotted horizontal line in Fig.~\ref{fig:New2}B).
For a low interaction strength Hertzian potential, this threshold is actually never exceeded and the system remains a viscoelastic liquid at all generalized area fractions.
This case corresponds to the triangles in Fig.~\ref{fig:New2}B.
In the case of the multi-Hertzian model (squares), within the range $0.95 < \zeta_{2D} < 1.27$, the diffusion coefficient decreases by four orders of magnitude, corresponding to the transition from a liquid to a glassy state.
With further increases in the $\zeta_{2D}$, $D$ exhibits slight oscillations: it increases between $1.25 < \zeta_{2D} < 2.75$, decreases for $2.75 < \zeta_{2D} < 4.0$ and rises again afterwards.
This qualitative trend appears to be general for Hertzian-like potentials. 
However, the Young modulus of the core term in the multi-Hertzian model can be tuned so that the absolute value of $D$ never exhibits reentrance to the fluid phase.
Such non-monotonic behavior is reported in the literature for Hertzian spheres in 3D \cite{Pam09} and may be attributed to complex motions of collective diffusion.

We thus stress that it is a property of the Hertzian model itself, since it occurs independently of the size polydispersity.
Since the multi-Hertzian model captures the fact observed experimentally that the monolayer of the ULC microgels remains solid-like for $\zeta_{2D}\gtrsim 1.25$, we select this model as the most appropriate to study the rheological properties under shear to compare with the interfacial rheology experiments discussed in the following. 

\begin{figure}
  \includegraphics[width=0.48\textwidth]{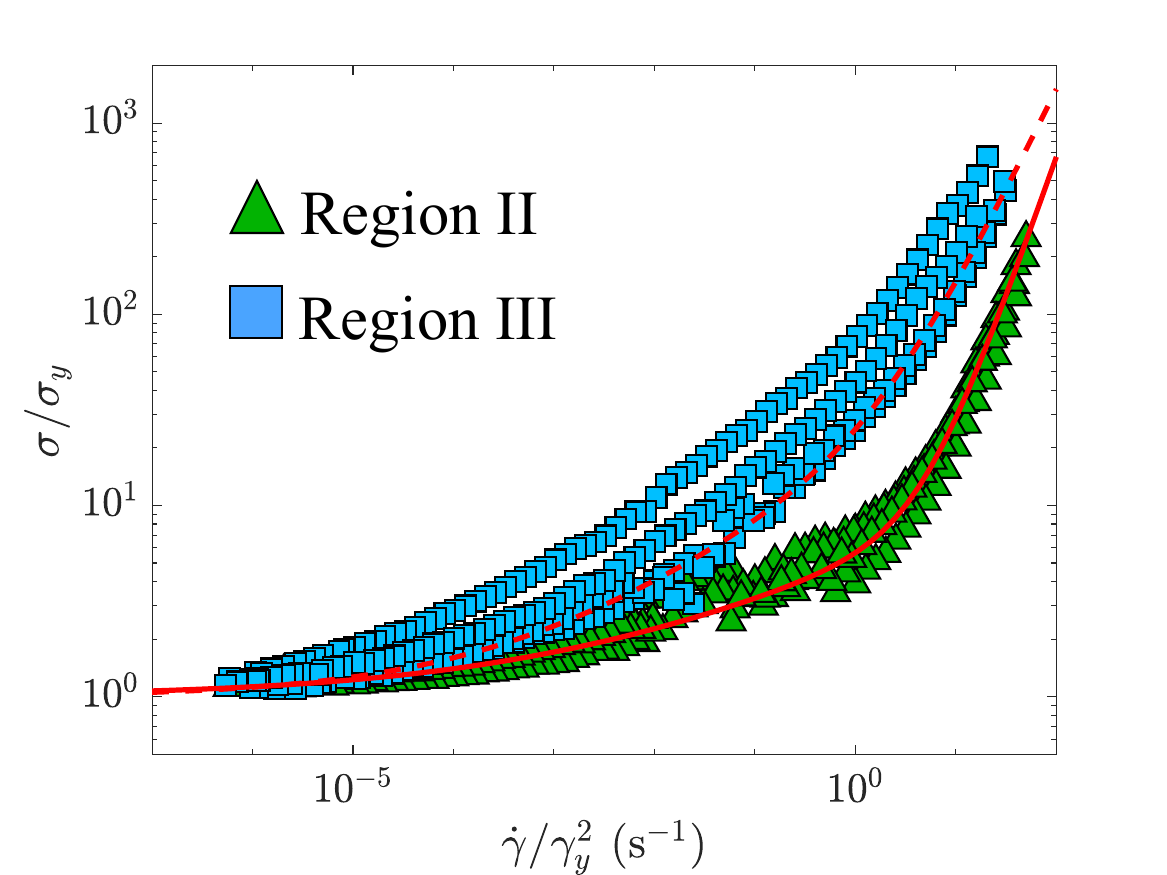}
  \centering
  \caption{Normalized flow curves of ULC microgel monolayers. Master curves obtained by the flow curves measured in the different compression stages where the stress $\sigma$ is normalized by the apparent yield stress $\sigma_y$ and the shear rate is normalized by $\gamma_y^2$, the squared value of the threshold strain amplitude between liquid-like and solid-like behavior. Solid and dashed lines correspond to $\sigma/\sigma_y = 1 + k(\gamma/\gamma_y^2)^u + k'(\gamma/\gamma_y^2)^p$ with $p \simeq 0.25$ and $u \simeq 1.5$ and $p \simeq 0.35$ and $u \simeq 1.1$, respectively.}
  \label{fig:ULC3FC}
\end{figure}

\subsection{Interfacial rheology experiments: linear shear}

We now shift our focus to the behavior of the monolayer under flow.
For a careful evaluation of the experimental data, and to compare the flow of the monolayer at different compression, one should use re-scaled variable.
Therefore, we normalize the stress under shear deformation in Fig.~\ref{fig:ULC3FC},
$\sigma$, by the apparent yield stress $\sigma_y$, obtained fitting the flow curves with a double power low \cite{Pel16, Sch23}.
The fits of the flow curves with Eq.~\ref{eq:HB_gen} are shown in Fig. S1.
Then, similarly to what was done in bulk \cite{Pel16} and at the interface \cite{Sch23}, we use the squared values of the threshold strain amplitude between solid-like and liquid-like behavior, as obtained from the amplitude sweeps shown in Fig. S2, to normalize the shear rate: $\dot\gamma/\gamma_y^2$.
With this re-scaling, the experimental flow curves for $\zeta_{2D} \lesssim 1.1$ (triangles) collapse onto a master curve with characteristic exponents in Eq.~\ref{eq:HB_gen}: $u = 1.50$ and $p = 0.25$. 
For $\zeta_{2D} \gtrsim 1.2$, the monolayer shows a second, distinct flow regime, characterised by different exponents, namely $u = 1.10$ and $p = 0.35$, in Eq.~\ref{eq:HB_gen}.
The fact that the last two curves for $\zeta_{2D}\gtrsim 2.10$ are a bit higher than the other data belonging to this second flow regime might be due to the difficulty in determining the value of $\gamma_y$ at such high concentration of the monolayer as discussed in the supplementary text. 
The fits of the experimental data and the corresponding fit parameters as a function of $\zeta_{2D}$  are reported for completeness in Fig. S1.

\begin{figure}
  \includegraphics[width=0.48\textwidth]{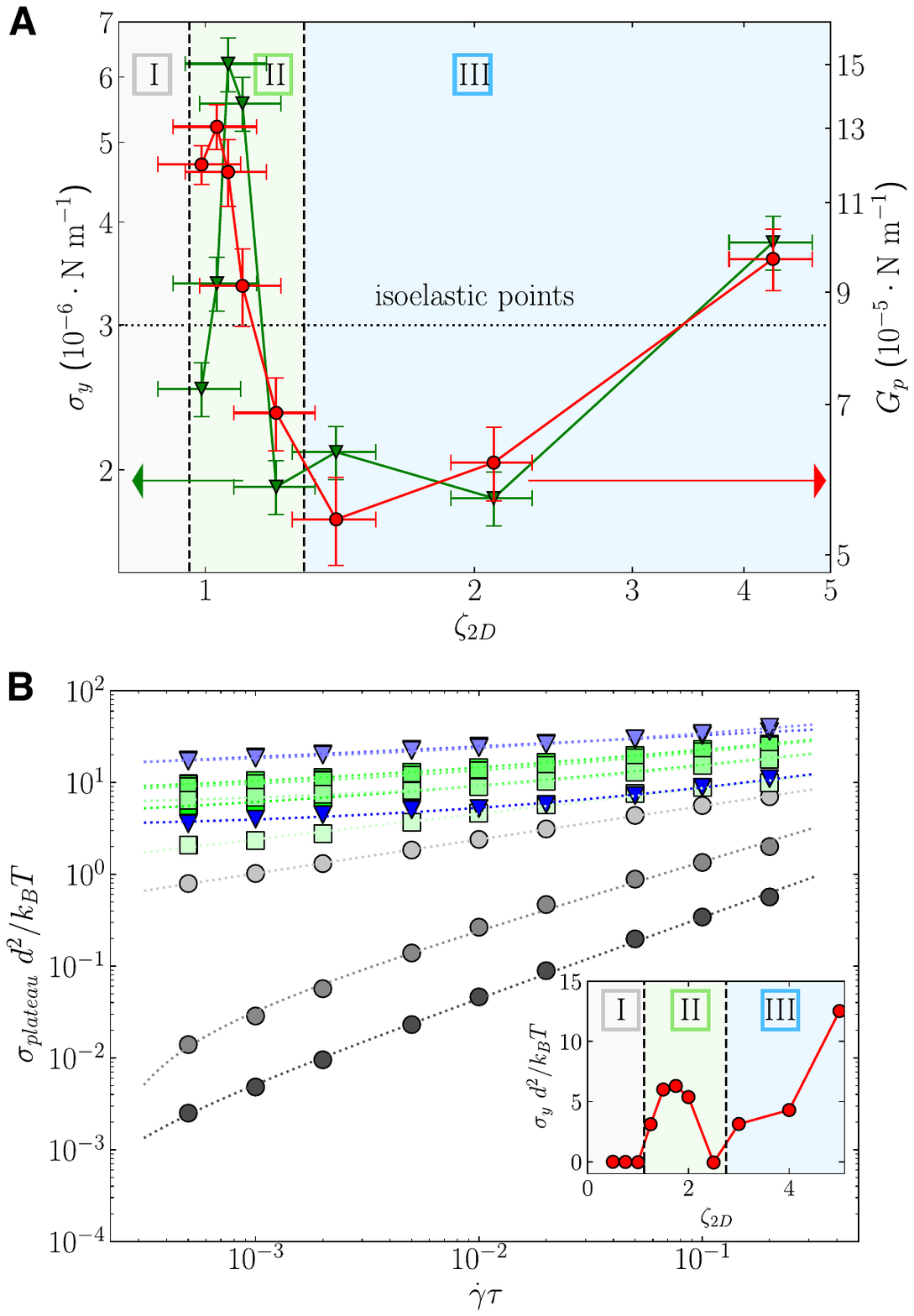}
  \centering
  \caption{\textbf{Elastic moduli of ULC microgel monolayers and simulated flow curves of multi-Hertzian discs.} (\textbf{A}) Values of the the apparent yield stress $\sigma_y$ (triangles) and of the plateau in the elastic modulus, $G_p$ (circles) as a function of the generalized area fraction. The arrows indicate the relevant y-axes for $\sigma_y$ (left) and $G_p$ (right). The horizontal dotted line indicates a representative \emph{isoelastic} point. (\textbf{B}) Simulated plateau modulus for generalized area fractions ranging between 0.5 - 5.0, extracted from stress strain curves. Circles refer to systems within region I, squares to region II and triangles to region III. The data (symobls) are fitted using a single power-law Herschel-Bulkley model scaling law (dotted line). Inset: The yield stress, $\sigma_y$, of the Herschel-Bulkley model, which extrapolates the behavior to zero shear. A value of $\sigma_y = 0$ signifies fluid behavior, whereas any nonzero value indicates glassy behavior in the system. The different regimes of the compression isotherms in Fig.~\ref{fig:ULC3Gr} are indicated by different colors.}
  \label{fig:New4}
\end{figure}

The presence of a master behavior for the flow curves is similar to what previously observed for the monolayer of harder microgels (5~mol\% crosslinker concentration with a bulk modulus of $\simeq 300$~kPa, i.e.~300 times larger than the one of ULC microgels \cite{Hou22}).
Indeed, even for hard microgels, depending on the monolayer compression, two distinct flow regimes have been identified \cite{Sch23}.
However, for harder microgels, additional regimes are present, and at high $\zeta_{2D}$, there is a superposition of the flow curves in regimes II and IV \cite{Sch23}.
This behavior is not observed for ULC microgels, as regime IV does not occur without a stiff core region.
Therefore, ULC microgels studied here seem to flow similarly to hard microgels in regimes II and III, suggesting a generality of the present findings, independently of particle softness and architecture.
The presence of different flow regimes has been explained since the microgels can be deformed both laterally and orthogonal to the interface \cite{Sch23, Ger24}.
As a result of this deformability, above a certain $\zeta_{2D}$, the viscous dissipation associated with the compression of the microgel volume is reduced setting up a different flow regime.

To support the fact that the deformability of the microgels affects how the energy is stored within the monolayer, the triangles in Fig.~\ref{fig:New4}A show the course of the apparent yield stress $\sigma_y$, obtained from fitting the flow curves in Fig. S1 with Eq.~\ref{eq:HB_gen}.
Clearly, $\sigma_y$ shows a non-monotonic behavior with increasing $\zeta_{2D}$ at first it increases, then decreases and finally increases again.
Remarkably, the monolayer shows the same values of $\sigma_y$ at very different $\zeta_{2D}$.
For instance, looking at the horizontal line in Fig.~\ref{fig:New4}A one can see that the point at $\zeta_{2D}\simeq 1.0$, $\simeq 1.15$ and $\simeq 3.2$ have a $\sigma_y \simeq 3\cdot10^{-6}$~Nm$^{-1}$.
This means that the same force has to be applied at these three very different concentrations to initiate the flow.
We define these points as \emph{isoelastic}.
A single cause of isoelasticity is difficult to identify.
It is also observed for conventional microgels ~\cite{Tat23,Sch23} at interfaces and related to the capability of the microgels to dissipate the energy of the lateral compression by deforming and protruding more in the water subphase as recently shown by neutron reflectometry and computer simulations \cite{Ger24}.
At the maximal yield stress, the interface is characterized by microgels that are deformed and interpenetrate each other, thereby covering the entire interfacial region.
For a further increase in generalized area fraction, the microgels necessarily shrink and deform even more.
This rearrangement lowers the force necessary to shear the interface.
Elucidating the structural changes might warrant further investigation using rheology and scattering techniques to directly measure the sheared structure \textit{in-situ}.

We now verify this behavior by performing non-equilibrium simulations of multi-Hertzian particles.
In a similar fashion to experiments, an external shear is applied and the resulting shear stress is computed (see Eq.~\ref{eq:IK_pressure}).
Flow curves, where the system is exposed to a linear shear with constant shear rate, are reported as symbols in Fig.~\ref{fig:New4}B.
All data can be well fitted by a standard Herschel-Bulkley scaling law i.e.~using Eq.~\ref{eq:HB_gen} with $k' = 0$, which are depicted as dotted lines.
This may be due to the fact that the simulated shear rate window is much smaller than the experimental one, for which two power-laws are necessary to fit the data.
Furthermore, the simulation does not consider any internal degrees of freedom of the network.
The values of $\sigma_y$ obtained fitting the simulated flow curves in the inset of Fig.~\ref{fig:New4}B, provide information about the yielding behavior of the system.
When $\sigma_y = 0$, the system is considered to be a fluid, as observed for all investigated state point in region I, which correspond to a generalized area fraction $\leq 1.0$.
These points were not accessible experimentally since the measured torque was below the rheometer accuracy.  
An increase in $\zeta_{2D}>1$ leads to finite values of $\sigma_y$, indicating that the system is a solid, namely a glass due to the lack of long-range order.
We also find that, consistently the with experimental data (triangles) in Fig.~\ref{fig:New4}A, the stress under viscous flow shows a non-monotonic behavior: it reaches a local maximum around $\zeta_{2D} \approx 1.75$  and then decreases with increasing concentration.
Therefore, also the simulations confirm the presence of \emph{isoelastic} points and the qualitative trend observed in experiments is fully reproduced by an effective pair potential treatment.
However, the precise $\zeta_{2D}$ at which \emph{isoelastic} points are observed and where the non-monotonic features arise differ from experimental ones due to the simplicity of the model.

\subsection{Interfacial rheology experiments: oscillatory shear}

Next, we focus on the results obtained from oscillatory frequency sweeps in the linear visco-elastic regime (Figs. S2 and S3). 
Despite the low values of the measured elastic ($G^{\prime}$) and viscous ($G^{\prime\prime}$) moduli, we observe that $G^{\prime}>G^{\prime\prime}$ for all investigated $\zeta_{2D}$, i.e.~the monolayer can be considered as a solid.
Since a clear minimum in $G^{\prime\prime}$ is missing, the mean and standard deviation of $G^{\prime}$ in each measurements are used as values and errors of $G_p$.
We thus report the plateau of the elastic modulus $G_p$ as a function of $\zeta_{2D}$ (circles) in Fig.~\ref{fig:New4}A.

It can be observed that $G_p$ has a non-monotonic dependence on the generalized area fraction, similarly to the yield stress reported in Fig.~\ref{fig:New4}A (triangles).
Again, this is in contrast with what reported in bulk for both hard \cite{Con19, Pel16, bergman2025free} and ULC microgels \cite{Sco20_flow, Sco19}.
Indeed, for those systems, $G_p$ was found to keep increasing with $\zeta_{2D}$, eventually reaching a plateau in the jammed state.
However, the results are qualitatively similar to what reported for harder microgels at the interface~\cite{Tat23,Sch23}.

When we compare the evolution of $G_p$ with the compression isotherms in Fig.~\ref{fig:ULC3Gr}(o), measured for the same monolayer of microgels at the oil-water interface, we see that the variation of the elastic properties are clearly linked to the variation of $\pi$.
When $\pi$ is roughly constant with $\zeta_{2D}$, namely in regions I and III, $G_p$ increases. In contrast, when $\pi$ steeply increases in region II, between $1.00 \lesssim \zeta_{2D} \lesssim 1.3$, $G_p$ significantly decreases.

Hence, when $\pi$ is constant or increases slightly, the system increasingly stores energy, because of the accumulation of the microgels within the available interface, which decreases due to the movement of the barriers of the trough.
However, the sharp growth observed in region II determines a strong compression of the monolayer, but since ULC microgels are highly deformable, they can accommodate more and more polymer chains onto the interface and protrude in the water subphase \cite{Ger24}.
Therefore, $G_p$ decreases because of the energy dissipation associated to the compression of the microgels. 

At high $\zeta_{2D}$, the evolution of $\pi$ is characterized by the absence of a second, sharp increase and consequently by a rather flat behavior of $G_p$. 
As mentioned before, this is a peculiarity of ULC microgels, which have an extremely homogeneous polymer distribution within their volume, in contrast to what was observed in the compression isotherms of more crosslinked microgels. 
For the latter, at some point the compression of the core enters into play, leading to a second increase of $\pi$ \cite{Rey16, Sco19, Gei14} and of $G_p$\cite{Sch23}.

The fact that the energy stored in the monolayer does not strongly increase  with increasing $\zeta_{2D}$ is consistent with the course of the apparent yield stress (see triangles in Fig.~\ref{fig:New4}).
Similarly to $G_p$, also $\sigma_y$ spans roughly one order of magnitude upon compression, associated to a non-monotonic behavior.
Importantly, at high compressions ($\zeta_{2D} \simeq 5$), $\sigma_y$ is measured to be smaller than the its maximum value in region II ($\zeta_{2D} \simeq 1$).
This is similar to what observed for $G_p$ which also, at high generalized area fractions, does not reach values as large as in region II.
Even for $G_p$, the non-monotonic course originates \emph{isoelastic} points, highlighted by the horizontal line in Fig.~\ref{fig:New4}A.
In this case, we see that the monolayer stores the same energy ($\simeq 8.25\cdot10^{-5}$~Nm$^{-1}$) at very different $\zeta_{2D}$. 
This peculiarity of microgels at interfaces, also shared with harder microgels~\cite{Sch23}, is similar to what previously observed for attractive colloidal glasses~\cite{Zac01,Pha08} undergoing a reentrant melting. 
While in that case, this was originated by the competition between two different mechanisms of glass formation, excluded volume and short-range attraction, in the present systems it naturally stems from the soft, Hertzian nature of the particles.

\begin{figure}
  \includegraphics[width=0.48\textwidth]{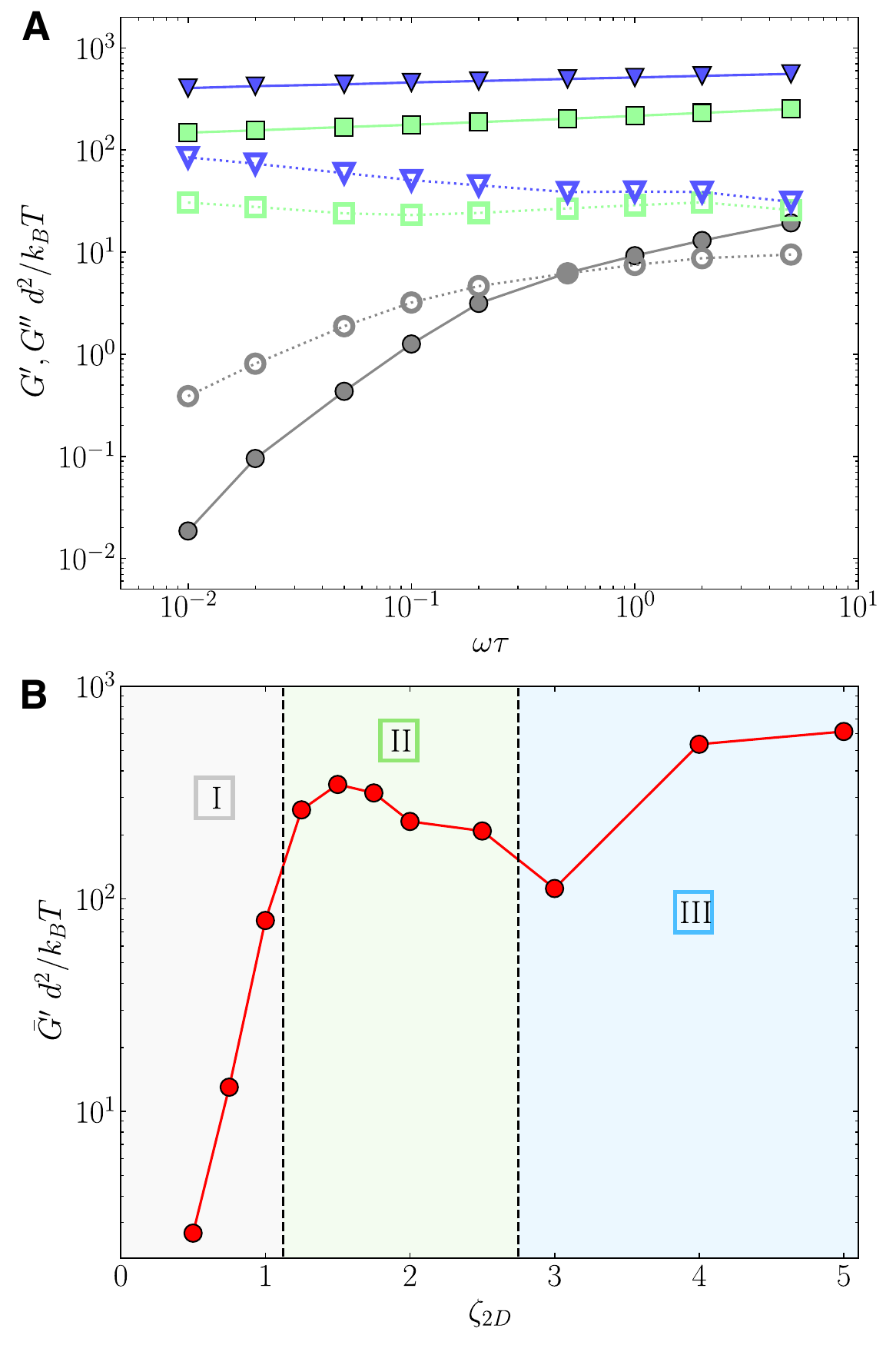}
  \centering
  \caption{\textbf{Oscillatory shear simulations of multi-Hertzian discs.} (\textbf{A}) Exemplary simulated storage (solid symbols) and loss modulus (hollow symbols) as a function of angular frequency for generalized area fractions of 0.75 (circles), 2.00 (squares) and 4.00 (triangles). The liquid phase in region I can be identified by a higher loss than storage modulus at low angular frequencies, while regions II and III are all glasses with higher storage modulus at all angular frequencies. (\textbf{B}) Averaged storage modulus as a function of $\zeta_{2D}$, showcasing the oscillating behavior seen experimentally. The different regimes of the compression isotherms in Fig.~\ref{fig:ULC3Gr} are indicated by different colors.}
  \label{fig:SIM_OS}
\end{figure}

To compare with experiments, simulated frequency sweeps are  addressed with a strain amplitude of $\gamma_0 = 3~\%$, ensuring a viscoelastic response in the linear regime.
In Fig.~\ref{fig:SIM_OS}A, one representative storage (solid symbols) and loss (empty symbols) modulus for each regime are plotted as a function of angular frequency.
An overview of all the data is given in Fig. S4.
Systems in the first regime (circles), exhibit a loss modulus higher than the storage modulus in the low-frequency region, confirming they behave as viscoelastic liquids.
For these concentrations, we do not have an experimental counterpart because the torque measured for the monolayer before microgels make contact was below the instrument sensitivity.

All denser systems in regime II (squares) and III (triangles) exhibit solid behavior, as the storage modulus exceeds the loss modulus across the entire frequency regime by about half an order of magnitude.
We emphasize that remaining in a solid state at high generalized area fractions is only achievable due to the use of a multi-Hertzian potential rather than with a pure Hertzian \cite{Sch23}. 
The storage modulus averaged over the simulated frequency range $\bar{G}'$ is shown by the symbols in Fig.~\ref{fig:SIM_OS}B as a function of $\zeta_{2D}$.
As expected, $\bar{G}'$ behaves roughly oppositely to the diffusion coefficient (Fig.~\ref{fig:New2}B), with a higher modulus corresponding to slower dynamics. 
Importantly, the storage modulus exhibits a qualitatively similar trend to the experiments, with an initial increase while the solid is formed, followed by a subsequent decrease and then a final, moderate increase (circles in Fig.~\ref{fig:New4}A). 
Furthermore, \emph{isoelastic} points are also observed in the oscillatory shear simulations.

The plot in Fig.~\ref{fig:SIM_OS}B is reported on a logarithmic scale to highlight the onset of solid behavior at low $\zeta_{2D}$, that could not be detected in the experiments. 
However, the simplicity of the model leads to some quantitative differences with experiments.
First of all, the location of regions II and III are slightly shifted with respect to the experimental ones, but they are consistent between linear and oscillatory shear simulations. 
Second, at very large values of $\zeta_{2D}$, the values of the moduli in the simulations slightly exceed those in region II, again due to the several approximations contained in the multi-Hertzian treatment. 
Nevertheless, we stress that this potential allows us to obtain a solid-like behavior in the entire range of $\zeta_{2D}$ above the onset of glassiness. 
This was not possible with a simple Hertzian and even with the square-shoulder Hertzian used in Ref.~\cite{Sch23} for harder microgels, so that we suggest that also in that case a multi-Hertzian model should be used plus a square-shoulder (to mimic the additional incompressible core of crosslinked microgel) in order to avoid the melting at high $\zeta_{2D}$. 
This will be explored in future work.

\subsection{Conclusion}

The present study investigates how monolayers of ultra-low crosslinked microgels interact at the interface in equilibrium and how they respond to external shear.
These microgels are even softer than regular microgels \cite{Hou22, Hof24}, since they can be continuously compressed up to very high generalized area fractions.
The observed behavior can be modelled using a multi-Hertzian potential \cite{bergman2018new}, where one Hertzian describes the microgel interaction for low deformation and additional Hertzian terms models the increase in stiffness upon compression.
This increased stiffness may also correlate with a thickening of the monolayer \cite{Ger24}.
Under shear, two distinct flow regimes can be identified.
At intermediate values of $\zeta_{2D}$, both the apparent yield stress and the storage modulus decrease.
The more easily compressible regions of the microgel monolayers are impacted first and the effective size of some microgels decreases.
Increasing $\zeta_{2D}$ further, the sheared monolayer is relatively homogeneous and the storage modulus increases again.
The non-monotonicity of both the yield stress and of the plateau modulus originates so-called \emph{isoelastic} points: at very different $\zeta_{2D}$ the monolayer presents comparable elastic properties.
The presence of these \emph{isoelastic} points, and more in general the non-monotonic course of both the yield stress and of the plateau in the elastic modulus, can be seen as the experimental actuation of the reentrant liquid phase expected for particles interacting with soft pair potentials.
It is very interesting to note that soft repulsion, here in the form of our simple Hertz-style potential but also present for many soft matter systems ranging from proteins to start polymers, is sufficient to yield this complex trend of elasticity.

Noteworthy, a similar behavior has been reported for harder microgels \cite{Sch23}.
Therefore, it seems to be connected to the softness of the interaction, independently on how it manifests.
However, differently from harder microgels, an incompressible core is not needed in the potential to model the behavior up to the very high generalized area fractions probed here. 
We stress that we reached the maximum possible compression of the monolayer, above which it is found to buckle or multilayers form. 
Therefore, this study demonstrates the very soft nature of the ULC microgels, which do not become incompressible even up to the maximum possible deformation. We expect this feature to also hold in bulk, opening the way to study theoretically the glass transition of these very soft objects. 
Indeed, previous studies have shown that the elasticity of soft colloidal polymers can influence the glass transition in bulk systems \cite{zou2024tunable}.

Finally, the fact that master curves can be identified also for ULC microgels indicates that this property is common to compressible spheres with bulk moduli from few to hundreds kPa \cite{Hou22}.
More studies on the flow properties of other synthetic colloids at the interface such as star polymers, DNA coated colloids, block co-polymers, and single chain nanoparticles, but also on bio-relevant macromolecules such as proteins and antibodies forming soft glasses \cite{Woo23}, will be needed in the future to understand whether these observations are universally related to the softness or are more specific of microgel systems.
More in general, a better understanding and modeling of the interaction between soft colloids at the interface can pave the way to a more rigorous description of biological interfaces.
For instance, a cellular membrane is composed of different type of soft colloids, e.g.~glycerophospholipids with  embedded different transmembrane proteins.
Each of these components posses its own intrinsic softness.
It is known that, depending on the particular biological activity, the cell membrane goes trough cycles of stiffening and softening \cite{wu2018comparison, komaragiri2024cell} which has to be related to the interaction of the different colloids and macromolecules composing the membrane.
Therefore, on one hand our theoretical understanding of these interactions and on the other the application of interfacial rheology can be pivotal in further advance our understating of these more complicated systems.
.\\

\noindent\textbf{Experimental and simulation section}\\
\noindent\textbf{Microgel synthesis} The microgels used in this study are identical to those described in Refs.~\cite{Sco19b, Hou22} and were synthesized through conventional precipitation polymerization \cite{Pel86}.
In summary, 8.4870~g of NIPAM, 0.6090~g of BIS, and 0.0560~g of SDS were dissolved in 495~mL of filtered double-distilled water. 
The solution was heated to 60~$^\circ$C and degassed by purging with nitrogen for one hour being continuously stirred at 300~rpm. 
The initiator solution, consisting of 0.2108~g of potassium persulfate (KPS) was diluted in 5~mL of filtered double-distilled water and degassed separately. 
Subsequently, the initiator was introduced rapidly into the monomer solution to trigger the polymerisation. 
After 4~hours, the reaction was terminated by lowering the temperature to ambient conditions. Finally, the sample was then purified using ultra-centrifugation at 30,000~rpm, followed by lyophilization for storage.\\

\noindent\textbf{Atomic Force Microscopy} ULC microgel suspensions were prepared according to Ref.~\cite{Sco19, Boc23, Sch23, Ger24} at a concentration of 0.25 wt~\% by redispersing lyophilized microgel powder in ultrapure water with 18.2~M$\Omega$cm$^{-1}$ resistivity. 
All suspensions were left on a rolling mixer overnight for equilibration.
Afterwards, microgel monolayers have been deposited at varying compression using a Langmuir-Blodgett through with rectangular geometry.
The height profile of the dried monolayer was measured at the solid-air interface ex situ in tapping mode with  OTESPA tip with a resonance frequency of 300 kHz, a nominal spring constant of 26 Nm$^{-1}$ of the cantilever and a nominal tip radius of $<7$~nm (Opus by Micromasch, Germany).
From the AFM images, the generalized area fraction is evaluated as:

\begin{equation}
    \zeta_{2D} = \frac{N_p A_p}{A_{tot}}\,\, ,
    \label{eq:z2D}
\end{equation}

\noindent where $A_{tot}$ is the total area of the image, $N_p$ is the number of microgels and $A_{p} = \pi R_{2D}^2$ is the area of an individual microgel in the dilute state.
Here, $R_{2D}$ is estimated as an average radius of over $220$ adsorbed ULC microgels at the interface before contact.
The radius distribution (compare \ref{fig:R_dist}) is fitted using a Gaussian, yielding $R_{2D} = 323\pm33$~nm.
When computing $\zeta_{2D}$, we consider $R_{2D}$ to be the primary source of error.
Its uncertainty is estimated using Gaussian error propagation based on the standard deviation of $R_{2D}$.\\

\noindent\textbf{Interfacial Rheology} Shear rheology measurements of ultra low crosslinked microgels adsorbed to an oil-water interface were performed.
As water phase, ultrapure water with a resistivity of  18.2~M$\Omega cm^{-1}$ was chosen and as oil phase we used Decane ($\leq 94$\%, Merck KGaA, Germany) filtered three times over basic aluminum oxide (90 standardized, Merck KGaA, Germany).
The apparatus consist of a Discovery HR-3 rheometer (TA Instruments Inc., USA) with a custom-made double wall-ring accessory combined with a purpose-built Langmuir trough\cite{Van10, Tat23}.
Using external barriers, the generalized area fraction can be controlled by the trough area.
The surface pressure is measured via a platinum Wilhelmy plate (KN 0002, KSV NIMA/Biolin Scientific Oy, Finland) with a perimeter of 39.24~mm.
Our setup and sample preparation is explained in more detail elsewhere\cite{Sch23}.
With the equilibrated monolayer at various generalized area fractions, we perform linear shear measurements to extract flow curves and oscillatory shear measurements in the form of amplitude and frequency sweep measurements.
The data relative to the flow curves in Fig.~\ref{fig:ULC3FC} are fitted using a linear combination of two power laws \cite{Pel16, Sch23}: 
\begin{equation}\label{eq:HB_gen}
    \sigma\left(\dot{\gamma}\right) = \sigma_{y}+k\dot{\gamma}^{u}+k'\dot{\gamma}^{p}
\end{equation}
where $k$, $k'$ are fitting parameters, $u$ and $p \in\mathbb{R}$.\\ 

\noindent\textbf{Effective potential} To model ULC microgels at the interface, we use a 2D Hertzian model that has been recently shown to well describe the pair potential of two monomer-resolved microgels at the interface~\cite{Cam20}.
Furthermore, some of us recently used a modified version of this potential for a similar study, revolving around interfacial rheology of regularly crosslinked microgels \cite{Sch23}.
The 2D Hertzian reads as:
\begin{equation}
V_H (r) = \frac{\pi Y d^2 \left(1-r/d\right)^2}{2 \ln\left(\frac{2}{1-r/d}\right)}, \,\, 
\label{eq:2dhertzian}
 \end{equation}
where $Y$ is the 2D Young modulus and $d$ is the effective diameter of the discs and the base unit length of our Molecular Dynamics simulations.
In general, all results are given in reduced Lennard-Jones units, which consist of: unit mass $m$, energy $\epsilon = k_BT$, time $\tau = \sqrt{\frac{md^{2}}{\epsilon}}$ and temperature $T = \frac{\epsilon}{k_B}$, where $k_B$ is the Boltzmann constant.
The simulation units can afterwards be rescaled to the experimental system, mainly by assuming $d$ is the size of the experimental microgels.
Simulations are performed within the framework of LAMMPS \cite{LAMMPS} in the canonical ensemble, with the temperature being set to 1 through a Langevin thermostat with a damping parameter of $\xi = 100\,m/\tau$.
We time integrate with a timestep $\Delta t = 0.002\,\tau$.
To estimate a comparable interaction strength to the experimental system, equilibrium simulations of $N=10000$ particles are performed in a range of generalized area fractions from 0.5 to 8.5.
The generalized area fraction is calculated as  $\zeta_{2D}=\frac{\pi}{4N}\sum_i^N d_{i}^2$, since the discs are polydisperse.
To best reflect the experimental system, we use a discrete size distribution with 50 particle sizes, according to a gaussian distribution with a polydisersity of 10
\%.
Hence, polydispersity is considered by averaging the diameter of two interacting particles, $i$ and $j$, as $d_{ij} = (d_i + d_j)/2$.
We find that a good agreement with the experimental results is obtained for $Y=100\, \epsilon/d^2$, yielding a Hertzian strength $\sim 226 \epsilon$, lower than the one estimated for hard microgels in Ref.~\cite{Sch23}.
To avoid system melting with increased concentration, we adopt a multi-Hertzian approach:

\begin{equation}
    V_{MH}(r) = \sum_i V_{H}(r,d_i,Y_i)
    \label{eq:MH}
\end{equation}

\noindent where an additional Hertzian core with higher Young's modulus and lower effective radii is considered. Eq.~\ref{eq:MH} is a sum of multiple Hertzian terms with different effective sizes, $d_i$, and moduli, $Y_i$.
For interactions between regions with different Young's modulus, the geometric mean of the two moduli describes the effective interaction.
Effectively, we add two additional Hertzian interaction terms: one between the stiff core region and the microgel, and another between two stiff core regions.
However, these additional terms do not describe an incompressible core interaction in the middle, but take into account the increase of the internal elasticity of the microgel toward the center, that is probed upon compression of the monolayer.
We find that adding a core contribution with  $Y_{core}=270 \epsilon/d^2$ and size $d_{core} = 0.9d$ is just enough to prevent the onset of a reentrant liquid phase without affecting the behavior of the system for low $\zeta_{2D}$.
This core-core term amounts to an additional Hertzian strength of $\approx 496\epsilon$, while the core-shell term, defined by $\sqrt{Y\cdot Y_{core}} \approx 164\epsilon/d^2$ and $\frac{1}{2}(d + d_{core}) = 0.95d$, adds approximately $335\epsilon$.\\

\noindent\textbf{Modeling Mechanical Shear} Rheological properties are investigated by imposing an artificial shear along the $xy$-plane, where $x$ is the direction of the flow and $y$ the one of the gradient.
Shear rates, $\dot{\gamma}$, are chosen in a range of $5\cdot10^{-3}\tau^{-1}$ to $5\cdot10^{-1}\tau^{-1}$.
The resulting shear stress, $\sigma_{xy}$, is calculated via the Irving–Kirkwood expression
\begin{equation}
    \sigma_{xy} = \frac{1}{A} \left( \sum_{i=1}^{N}\left[m_{i}v_{i,x}v_{i,y} + \sum_{j>i}^{N}r_{ij,x}f_{ij,y} \right]\right)\,,
    \label{eq:IK_pressure}
\end{equation}
where $A$ is the total area,
$v_{x}$ is the $x$ component of the velocity and $f_y$ is the $y$ component of the force.
With increasing strain, $\sigma_{xy}$ initially increases, then it overcomes a maxima that represents the yielding and finally reaches a steady state for large shear, which is the plateau modulus, $\sigma_{plateau}$.
By plotting the plateau modulus as a function of shear rate, flow curves are obtained.\\

We also apply an oscillatory shear rate, where the shear has the function of a sine wave, with a strain amplitude of $\gamma_{0} = 3$\%, to stay in the linear visco-elastic regime.
The angular frequency, $\omega$, is then varied from $1\cdot10^{-2}\tau^{-1}$ to $5\cdot10^{0}\tau^{-1}$.
The resulting stress signal is a sine wave as well, so the storage, $G'$, and loss modulus, $G''$ can be calculated as:
\begin{align}
    G'(\omega) &= Re\left( \frac{\bar{\sigma}}{\bar{\gamma}} \right) \\
    G''(\omega) &= Im\left( \frac{\bar{\sigma}}{\bar{\gamma}} \right)
\end{align}
with the $\bar{\sigma}$ and $\bar{\gamma}$ being the Fourier transform of the stress and shear strain respectively.

\noindent\textbf{Data availability}\\
\noindent 
The data that support the findings of this study are available from the corresponding author upon reasonable request.

\bibliographystyle{apsrev4-1}
\bibliography{References}

\begin{thebibliography}{50}%
\makeatletter
\providecommand \@ifxundefined [1]{%
 \@ifx{#1\undefined}
}%
\providecommand \@ifnum [1]{%
 \ifnum #1\expandafter \@firstoftwo
 \else \expandafter \@secondoftwo
 \fi
}%
\providecommand \@ifx [1]{%
 \ifx #1\expandafter \@firstoftwo
 \else \expandafter \@secondoftwo
 \fi
}%
\providecommand \natexlab [1]{#1}%
\providecommand \enquote  [1]{``#1''}%
\providecommand \bibnamefont  [1]{#1}%
\providecommand \bibfnamefont [1]{#1}%
\providecommand \citenamefont [1]{#1}%
\providecommand \href@noop [0]{\@secondoftwo}%
\providecommand \href [0]{\begingroup \@sanitize@url \@href}%
\providecommand \@href[1]{\@@startlink{#1}\@@href}%
\providecommand \@@href[1]{\endgroup#1\@@endlink}%
\providecommand \@sanitize@url [0]{\catcode `\\12\catcode `\$12\catcode
  `\&12\catcode `\#12\catcode `\^12\catcode `\_12\catcode `\%12\relax}%
\providecommand \@@startlink[1]{}%
\providecommand \@@endlink[0]{}%
\providecommand \url  [0]{\begingroup\@sanitize@url \@url }%
\providecommand \@url [1]{\endgroup\@href {#1}{\urlprefix }}%
\providecommand \urlprefix  [0]{URL }%
\providecommand \Eprint [0]{\href }%
\providecommand \doibase [0]{http://dx.doi.org/}%
\providecommand \selectlanguage [0]{\@gobble}%
\providecommand \bibinfo  [0]{\@secondoftwo}%
\providecommand \bibfield  [0]{\@secondoftwo}%
\providecommand \translation [1]{[#1]}%
\providecommand \BibitemOpen [0]{}%
\providecommand \bibitemStop [0]{}%
\providecommand \bibitemNoStop [0]{.\EOS\space}%
\providecommand \EOS [0]{\spacefactor3000\relax}%
\providecommand \BibitemShut  [1]{\csname bibitem#1\endcsname}%
\let\auto@bib@innerbib\@empty
\bibitem [{\citenamefont {de~Gennes}(2005)}]{deG05}%
  \BibitemOpen
  \bibfield  {author} {\bibinfo {author} {\bibfnamefont {P.-G.}\ \bibnamefont
  {de~Gennes}},\ }\href@noop {} {\bibfield  {journal} {\bibinfo  {journal}
  {Soft Matter}\ }\textbf {\bibinfo {volume} {1}},\ \bibinfo {pages} {16}
  (\bibinfo {year} {2005})}\BibitemShut {NoStop}%
\bibitem [{\citenamefont {Segre}\ \emph {et~al.}(1995)\citenamefont {Segre},
  \citenamefont {Meeker}, \citenamefont {Pusey},\ and\ \citenamefont
  {Poon}}]{Seg95}%
  \BibitemOpen
  \bibfield  {author} {\bibinfo {author} {\bibfnamefont {P.}~\bibnamefont
  {Segre}}, \bibinfo {author} {\bibfnamefont {S.}~\bibnamefont {Meeker}},
  \bibinfo {author} {\bibfnamefont {P.}~\bibnamefont {Pusey}}, \ and\ \bibinfo
  {author} {\bibfnamefont {W.}~\bibnamefont {Poon}},\ }\href@noop {} {\bibfield
   {journal} {\bibinfo  {journal} {Physical review letters}\ }\textbf {\bibinfo
  {volume} {75}},\ \bibinfo {pages} {958} (\bibinfo {year} {1995})}\BibitemShut
  {NoStop}%
\bibitem [{\citenamefont {Van~Megen}\ and\ \citenamefont
  {Underwood}(1994)}]{van1994glass}%
  \BibitemOpen
  \bibfield  {author} {\bibinfo {author} {\bibfnamefont {W.}~\bibnamefont
  {Van~Megen}}\ and\ \bibinfo {author} {\bibfnamefont {S.}~\bibnamefont
  {Underwood}},\ }\href@noop {} {\bibfield  {journal} {\bibinfo  {journal}
  {Physical Review E}\ }\textbf {\bibinfo {volume} {49}},\ \bibinfo {pages}
  {4206} (\bibinfo {year} {1994})}\BibitemShut {NoStop}%
\bibitem [{\citenamefont {Vlassopoulos}\ and\ \citenamefont
  {Cloitre}(2014)}]{Vla14}%
  \BibitemOpen
  \bibfield  {author} {\bibinfo {author} {\bibfnamefont {D.}~\bibnamefont
  {Vlassopoulos}}\ and\ \bibinfo {author} {\bibfnamefont {M.}~\bibnamefont
  {Cloitre}},\ }\href@noop {} {\bibfield  {journal} {\bibinfo  {journal}
  {Current opinion in colloid \& interface science}\ }\textbf {\bibinfo
  {volume} {19}},\ \bibinfo {pages} {561} (\bibinfo {year} {2014})}\BibitemShut
  {NoStop}%
\bibitem [{\citenamefont {Scotti}\ \emph {et~al.}(2022)\citenamefont {Scotti},
  \citenamefont {Schulte}, \citenamefont {Lopez}, \citenamefont {Crassous},
  \citenamefont {Bochenek},\ and\ \citenamefont {Richtering}}]{Sco22_review}%
  \BibitemOpen
  \bibfield  {author} {\bibinfo {author} {\bibfnamefont {A.}~\bibnamefont
  {Scotti}}, \bibinfo {author} {\bibfnamefont {M.~F.}\ \bibnamefont {Schulte}},
  \bibinfo {author} {\bibfnamefont {C.~G.}\ \bibnamefont {Lopez}}, \bibinfo
  {author} {\bibfnamefont {J.~J.}\ \bibnamefont {Crassous}}, \bibinfo {author}
  {\bibfnamefont {S.}~\bibnamefont {Bochenek}}, \ and\ \bibinfo {author}
  {\bibfnamefont {W.}~\bibnamefont {Richtering}},\ }\href {\doibase
  10.1021/acs.chemrev.2c00035} {\bibfield  {journal} {\bibinfo  {journal}
  {Chemical Reviews}\ }\textbf {\bibinfo {volume} {122}},\ \bibinfo {pages}
  {11675} (\bibinfo {year} {2022})}\BibitemShut {NoStop}%
\bibitem [{\citenamefont {Schmidt}\ \emph {et~al.}(2023)\citenamefont
  {Schmidt}, \citenamefont {Ruiz-Franco}, \citenamefont {Bochenek},
  \citenamefont {Camerin}, \citenamefont {Zaccarelli},\ and\ \citenamefont
  {Scotti}}]{Sch23}%
  \BibitemOpen
  \bibfield  {author} {\bibinfo {author} {\bibfnamefont {M.~M.}\ \bibnamefont
  {Schmidt}}, \bibinfo {author} {\bibfnamefont {J.}~\bibnamefont
  {Ruiz-Franco}}, \bibinfo {author} {\bibfnamefont {S.}~\bibnamefont
  {Bochenek}}, \bibinfo {author} {\bibfnamefont {F.}~\bibnamefont {Camerin}},
  \bibinfo {author} {\bibfnamefont {E.}~\bibnamefont {Zaccarelli}}, \ and\
  \bibinfo {author} {\bibfnamefont {A.}~\bibnamefont {Scotti}},\ }\href@noop {}
  {\bibfield  {journal} {\bibinfo  {journal} {Physical Review Letters}\
  }\textbf {\bibinfo {volume} {131}},\ \bibinfo {pages} {258202} (\bibinfo
  {year} {2023})}\BibitemShut {NoStop}%
\bibitem [{\citenamefont {Erwin}\ \emph {et~al.}(2010)\citenamefont {Erwin},
  \citenamefont {Cloitre}, \citenamefont {Gauthier},\ and\ \citenamefont
  {Vlassopoulos}}]{Erw10}%
  \BibitemOpen
  \bibfield  {author} {\bibinfo {author} {\bibfnamefont {B.~M.}\ \bibnamefont
  {Erwin}}, \bibinfo {author} {\bibfnamefont {M.}~\bibnamefont {Cloitre}},
  \bibinfo {author} {\bibfnamefont {M.}~\bibnamefont {Gauthier}}, \ and\
  \bibinfo {author} {\bibfnamefont {D.}~\bibnamefont {Vlassopoulos}},\
  }\href@noop {} {\bibfield  {journal} {\bibinfo  {journal} {Soft Matter}\
  }\textbf {\bibinfo {volume} {6}},\ \bibinfo {pages} {2825} (\bibinfo {year}
  {2010})}\BibitemShut {NoStop}%
\bibitem [{\citenamefont {Borrega}\ \emph {et~al.}(1999)\citenamefont
  {Borrega}, \citenamefont {Cloitre}, \citenamefont {Betremieux}, \citenamefont
  {Ernst},\ and\ \citenamefont {Leibler}}]{Bor99}%
  \BibitemOpen
  \bibfield  {author} {\bibinfo {author} {\bibfnamefont {R.}~\bibnamefont
  {Borrega}}, \bibinfo {author} {\bibfnamefont {M.}~\bibnamefont {Cloitre}},
  \bibinfo {author} {\bibfnamefont {I.}~\bibnamefont {Betremieux}}, \bibinfo
  {author} {\bibfnamefont {B.}~\bibnamefont {Ernst}}, \ and\ \bibinfo {author}
  {\bibfnamefont {L.}~\bibnamefont {Leibler}},\ }\href@noop {} {\bibfield
  {journal} {\bibinfo  {journal} {Europhysics Letters}\ }\textbf {\bibinfo
  {volume} {47}},\ \bibinfo {pages} {729} (\bibinfo {year} {1999})}\BibitemShut
  {NoStop}%
\bibitem [{\citenamefont {Scotti}\ \emph {et~al.}(2020)\citenamefont {Scotti},
  \citenamefont {Brugnoni}, \citenamefont {Lopez}, \citenamefont {Bochenek},
  \citenamefont {Crassous},\ and\ \citenamefont {Richtering}}]{Sco20_flow}%
  \BibitemOpen
  \bibfield  {author} {\bibinfo {author} {\bibfnamefont {A.}~\bibnamefont
  {Scotti}}, \bibinfo {author} {\bibfnamefont {M.}~\bibnamefont {Brugnoni}},
  \bibinfo {author} {\bibfnamefont {C.~G.}\ \bibnamefont {Lopez}}, \bibinfo
  {author} {\bibfnamefont {S.}~\bibnamefont {Bochenek}}, \bibinfo {author}
  {\bibfnamefont {J.~J.}\ \bibnamefont {Crassous}}, \ and\ \bibinfo {author}
  {\bibfnamefont {W.}~\bibnamefont {Richtering}},\ }\href@noop {} {\bibfield
  {journal} {\bibinfo  {journal} {Soft Matter}\ }\textbf {\bibinfo {volume}
  {16}},\ \bibinfo {pages} {668} (\bibinfo {year} {2020})}\BibitemShut
  {NoStop}%
\bibitem [{\citenamefont {Gao}\ and\ \citenamefont {Frisken}(2003)}]{Gao03}%
  \BibitemOpen
  \bibfield  {author} {\bibinfo {author} {\bibfnamefont {J.}~\bibnamefont
  {Gao}}\ and\ \bibinfo {author} {\bibfnamefont {B.~J.}\ \bibnamefont
  {Frisken}},\ }\href@noop {} {\bibfield  {journal} {\bibinfo  {journal}
  {Langmuir}\ }\textbf {\bibinfo {volume} {19}},\ \bibinfo {pages} {5212}
  (\bibinfo {year} {2003})}\BibitemShut {NoStop}%
\bibitem [{\citenamefont {Brugnoni}\ \emph {et~al.}(2019)\citenamefont
  {Brugnoni}, \citenamefont {Nickel}, \citenamefont {Kr{\"o}ger}, \citenamefont
  {Scotti}, \citenamefont {Pich}, \citenamefont {Leonhard},\ and\ \citenamefont
  {Richtering}}]{Bru19}%
  \BibitemOpen
  \bibfield  {author} {\bibinfo {author} {\bibfnamefont {M.}~\bibnamefont
  {Brugnoni}}, \bibinfo {author} {\bibfnamefont {A.~C.}\ \bibnamefont
  {Nickel}}, \bibinfo {author} {\bibfnamefont {L.~C.}\ \bibnamefont
  {Kr{\"o}ger}}, \bibinfo {author} {\bibfnamefont {A.}~\bibnamefont {Scotti}},
  \bibinfo {author} {\bibfnamefont {A.}~\bibnamefont {Pich}}, \bibinfo {author}
  {\bibfnamefont {K.}~\bibnamefont {Leonhard}}, \ and\ \bibinfo {author}
  {\bibfnamefont {W.}~\bibnamefont {Richtering}},\ }\href@noop {} {\bibfield
  {journal} {\bibinfo  {journal} {Polymer Chemistry}\ }\textbf {\bibinfo
  {volume} {10}},\ \bibinfo {pages} {2397} (\bibinfo {year}
  {2019})}\BibitemShut {NoStop}%
\bibitem [{\citenamefont {Houston}\ \emph {et~al.}(2022)\citenamefont
  {Houston}, \citenamefont {Fruhner}, \citenamefont {de~la Cotte},
  \citenamefont {Rojo~Gonz{\'a}lez}, \citenamefont {Petrunin}, \citenamefont
  {Gasser}, \citenamefont {Schweins}, \citenamefont {Allgaier}, \citenamefont
  {Richtering}, \citenamefont {Fernandez-Nieves} \emph {et~al.}}]{Hou22}%
  \BibitemOpen
  \bibfield  {author} {\bibinfo {author} {\bibfnamefont {J.~E.}\ \bibnamefont
  {Houston}}, \bibinfo {author} {\bibfnamefont {L.}~\bibnamefont {Fruhner}},
  \bibinfo {author} {\bibfnamefont {A.}~\bibnamefont {de~la Cotte}}, \bibinfo
  {author} {\bibfnamefont {J.}~\bibnamefont {Rojo~Gonz{\'a}lez}}, \bibinfo
  {author} {\bibfnamefont {A.~V.}\ \bibnamefont {Petrunin}}, \bibinfo {author}
  {\bibfnamefont {U.}~\bibnamefont {Gasser}}, \bibinfo {author} {\bibfnamefont
  {R.}~\bibnamefont {Schweins}}, \bibinfo {author} {\bibfnamefont
  {J.}~\bibnamefont {Allgaier}}, \bibinfo {author} {\bibfnamefont
  {W.}~\bibnamefont {Richtering}}, \bibinfo {author} {\bibfnamefont
  {A.}~\bibnamefont {Fernandez-Nieves}},  \emph {et~al.},\ }\href@noop {}
  {\bibfield  {journal} {\bibinfo  {journal} {Science Advances}\ }\textbf
  {\bibinfo {volume} {8}},\ \bibinfo {pages} {eabn6129} (\bibinfo {year}
  {2022})}\BibitemShut {NoStop}%
\bibitem [{\citenamefont {H{\"o}fken}\ \emph {et~al.}(2024)\citenamefont
  {H{\"o}fken}, \citenamefont {Gasser}, \citenamefont {Schneider},
  \citenamefont {Petrunin},\ and\ \citenamefont {Scotti}}]{Hof24}%
  \BibitemOpen
  \bibfield  {author} {\bibinfo {author} {\bibfnamefont {T.}~\bibnamefont
  {H{\"o}fken}}, \bibinfo {author} {\bibfnamefont {U.}~\bibnamefont {Gasser}},
  \bibinfo {author} {\bibfnamefont {S.}~\bibnamefont {Schneider}}, \bibinfo
  {author} {\bibfnamefont {A.~V.}\ \bibnamefont {Petrunin}}, \ and\ \bibinfo
  {author} {\bibfnamefont {A.}~\bibnamefont {Scotti}},\ }\href@noop {}
  {\bibfield  {journal} {\bibinfo  {journal} {Macromolecular rapid
  communications}\ ,\ \bibinfo {pages} {2400043}} (\bibinfo {year}
  {2024})}\BibitemShut {NoStop}%
\bibitem [{\citenamefont {Scotti}(2021)}]{Sco21}%
  \BibitemOpen
  \bibfield  {author} {\bibinfo {author} {\bibfnamefont {A.}~\bibnamefont
  {Scotti}},\ }\href@noop {} {\bibfield  {journal} {\bibinfo  {journal} {Soft
  Matter}\ }\textbf {\bibinfo {volume} {17}},\ \bibinfo {pages} {5548}
  (\bibinfo {year} {2021})}\BibitemShut {NoStop}%
\bibitem [{\citenamefont {Pusey}\ and\ \citenamefont
  {Van~Megen}(1986)}]{Pus86}%
  \BibitemOpen
  \bibfield  {author} {\bibinfo {author} {\bibfnamefont {P.~N.}\ \bibnamefont
  {Pusey}}\ and\ \bibinfo {author} {\bibfnamefont {W.}~\bibnamefont
  {Van~Megen}},\ }\href@noop {} {\bibfield  {journal} {\bibinfo  {journal}
  {Nature}\ }\textbf {\bibinfo {volume} {320}},\ \bibinfo {pages} {340}
  (\bibinfo {year} {1986})}\BibitemShut {NoStop}%
\bibitem [{\citenamefont {Scotti}\ \emph
  {et~al.}(2019{\natexlab{a}})\citenamefont {Scotti}, \citenamefont {Bochenek},
  \citenamefont {Brugnoni}, \citenamefont {Fernandez-Rodriguez}, \citenamefont
  {Schulte}, \citenamefont {Houston}, \citenamefont {Gelissen}, \citenamefont
  {Potemkin}, \citenamefont {Isa},\ and\ \citenamefont {Richtering}}]{Sco19}%
  \BibitemOpen
  \bibfield  {author} {\bibinfo {author} {\bibfnamefont {A.}~\bibnamefont
  {Scotti}}, \bibinfo {author} {\bibfnamefont {S.}~\bibnamefont {Bochenek}},
  \bibinfo {author} {\bibfnamefont {M.}~\bibnamefont {Brugnoni}}, \bibinfo
  {author} {\bibfnamefont {M.-A.}\ \bibnamefont {Fernandez-Rodriguez}},
  \bibinfo {author} {\bibfnamefont {M.~F.}\ \bibnamefont {Schulte}}, \bibinfo
  {author} {\bibfnamefont {J.}~\bibnamefont {Houston}}, \bibinfo {author}
  {\bibfnamefont {A.~P.}\ \bibnamefont {Gelissen}}, \bibinfo {author}
  {\bibfnamefont {I.~I.}\ \bibnamefont {Potemkin}}, \bibinfo {author}
  {\bibfnamefont {L.}~\bibnamefont {Isa}}, \ and\ \bibinfo {author}
  {\bibfnamefont {W.}~\bibnamefont {Richtering}},\ }\href@noop {} {\bibfield
  {journal} {\bibinfo  {journal} {Nature Communications}\ }\textbf {\bibinfo
  {volume} {10}},\ \bibinfo {pages} {1} (\bibinfo {year}
  {2019}{\natexlab{a}})}\BibitemShut {NoStop}%
\bibitem [{\citenamefont {Schulte}\ \emph {et~al.}(2021)\citenamefont
  {Schulte}, \citenamefont {Bochenek}, \citenamefont {Brugnoni}, \citenamefont
  {Scotti}, \citenamefont {Mourran},\ and\ \citenamefont {Richtering}}]{Sch21}%
  \BibitemOpen
  \bibfield  {author} {\bibinfo {author} {\bibfnamefont {M.~F.}\ \bibnamefont
  {Schulte}}, \bibinfo {author} {\bibfnamefont {S.}~\bibnamefont {Bochenek}},
  \bibinfo {author} {\bibfnamefont {M.}~\bibnamefont {Brugnoni}}, \bibinfo
  {author} {\bibfnamefont {A.}~\bibnamefont {Scotti}}, \bibinfo {author}
  {\bibfnamefont {A.}~\bibnamefont {Mourran}}, \ and\ \bibinfo {author}
  {\bibfnamefont {W.}~\bibnamefont {Richtering}},\ }\href {\doibase
  https://doi.org/10.1002/anie.202011615} {\bibfield  {journal} {\bibinfo
  {journal} {Angewandte Chemie International Edition}\ }\textbf {\bibinfo
  {volume} {60}},\ \bibinfo {pages} {2280} (\bibinfo {year}
  {2021})}\BibitemShut {NoStop}%
\bibitem [{\citenamefont {Rey}\ \emph {et~al.}(2023)\citenamefont {Rey},
  \citenamefont {Kolker}, \citenamefont {Richards}, \citenamefont {Malhotra},
  \citenamefont {Glen}, \citenamefont {Li}, \citenamefont {Laidlaw},
  \citenamefont {Renggli}, \citenamefont {Vermant}, \citenamefont {Schofield}
  \emph {et~al.}}]{Rey23}%
  \BibitemOpen
  \bibfield  {author} {\bibinfo {author} {\bibfnamefont {M.}~\bibnamefont
  {Rey}}, \bibinfo {author} {\bibfnamefont {J.}~\bibnamefont {Kolker}},
  \bibinfo {author} {\bibfnamefont {J.~A.}\ \bibnamefont {Richards}}, \bibinfo
  {author} {\bibfnamefont {I.}~\bibnamefont {Malhotra}}, \bibinfo {author}
  {\bibfnamefont {T.~S.}\ \bibnamefont {Glen}}, \bibinfo {author}
  {\bibfnamefont {N.~D.}\ \bibnamefont {Li}}, \bibinfo {author} {\bibfnamefont
  {F.~H.}\ \bibnamefont {Laidlaw}}, \bibinfo {author} {\bibfnamefont
  {D.}~\bibnamefont {Renggli}}, \bibinfo {author} {\bibfnamefont
  {J.}~\bibnamefont {Vermant}}, \bibinfo {author} {\bibfnamefont {A.~B.}\
  \bibnamefont {Schofield}},  \emph {et~al.},\ }\href@noop {} {\bibfield
  {journal} {\bibinfo  {journal} {Nature Communications}\ }\textbf {\bibinfo
  {volume} {14}},\ \bibinfo {pages} {6723} (\bibinfo {year}
  {2023})}\BibitemShut {NoStop}%
\bibitem [{\citenamefont {Bochenek}\ \emph {et~al.}(2022)\citenamefont
  {Bochenek}, \citenamefont {Camerin}, \citenamefont {Zaccarelli},
  \citenamefont {Maestro}, \citenamefont {Schmidt}, \citenamefont
  {Richtering},\ and\ \citenamefont {Scotti}}]{Boc22}%
  \BibitemOpen
  \bibfield  {author} {\bibinfo {author} {\bibfnamefont {S.}~\bibnamefont
  {Bochenek}}, \bibinfo {author} {\bibfnamefont {F.}~\bibnamefont {Camerin}},
  \bibinfo {author} {\bibfnamefont {E.}~\bibnamefont {Zaccarelli}}, \bibinfo
  {author} {\bibfnamefont {A.}~\bibnamefont {Maestro}}, \bibinfo {author}
  {\bibfnamefont {M.~M.}\ \bibnamefont {Schmidt}}, \bibinfo {author}
  {\bibfnamefont {W.}~\bibnamefont {Richtering}}, \ and\ \bibinfo {author}
  {\bibfnamefont {A.}~\bibnamefont {Scotti}},\ }\href@noop {} {\bibfield
  {journal} {\bibinfo  {journal} {Nature Communications}\ }\textbf {\bibinfo
  {volume} {13}},\ \bibinfo {pages} {3744} (\bibinfo {year}
  {2022})}\BibitemShut {NoStop}%
\bibitem [{\citenamefont {Petrunin}\ \emph {et~al.}(2023)\citenamefont
  {Petrunin}, \citenamefont {Bochenek}, \citenamefont {Richtering},\ and\
  \citenamefont {Scotti}}]{Pet23}%
  \BibitemOpen
  \bibfield  {author} {\bibinfo {author} {\bibfnamefont {A.~V.}\ \bibnamefont
  {Petrunin}}, \bibinfo {author} {\bibfnamefont {S.}~\bibnamefont {Bochenek}},
  \bibinfo {author} {\bibfnamefont {W.}~\bibnamefont {Richtering}}, \ and\
  \bibinfo {author} {\bibfnamefont {A.}~\bibnamefont {Scotti}},\ }\href@noop {}
  {\bibfield  {journal} {\bibinfo  {journal} {Physical Chemistry Chemical
  Physics}\ }\textbf {\bibinfo {volume} {25}},\ \bibinfo {pages} {2810}
  (\bibinfo {year} {2023})}\BibitemShut {NoStop}%
\bibitem [{\citenamefont {Gerelli}\ \emph {et~al.}(2024)\citenamefont
  {Gerelli}, \citenamefont {Camerin}, \citenamefont {Bochenek}, \citenamefont
  {Schmidt}, \citenamefont {Maestro}, \citenamefont {Richtering}, \citenamefont
  {Zaccarelli},\ and\ \citenamefont {Scotti}}]{Ger24}%
  \BibitemOpen
  \bibfield  {author} {\bibinfo {author} {\bibfnamefont {Y.}~\bibnamefont
  {Gerelli}}, \bibinfo {author} {\bibfnamefont {F.}~\bibnamefont {Camerin}},
  \bibinfo {author} {\bibfnamefont {S.}~\bibnamefont {Bochenek}}, \bibinfo
  {author} {\bibfnamefont {M.~M.}\ \bibnamefont {Schmidt}}, \bibinfo {author}
  {\bibfnamefont {A.}~\bibnamefont {Maestro}}, \bibinfo {author} {\bibfnamefont
  {W.}~\bibnamefont {Richtering}}, \bibinfo {author} {\bibfnamefont
  {E.}~\bibnamefont {Zaccarelli}}, \ and\ \bibinfo {author} {\bibfnamefont
  {A.}~\bibnamefont {Scotti}},\ }\href@noop {} {\bibfield  {journal} {\bibinfo
  {journal} {Soft Matter}\ }\textbf {\bibinfo {volume} {20}},\ \bibinfo {pages}
  {3653} (\bibinfo {year} {2024})}\BibitemShut {NoStop}%
\bibitem [{\citenamefont {Paloli}\ \emph {et~al.}(2013)\citenamefont {Paloli},
  \citenamefont {Mohanty}, \citenamefont {Crassous}, \citenamefont
  {Zaccarelli},\ and\ \citenamefont {Schurtenberger}}]{paloli2013fluid}%
  \BibitemOpen
  \bibfield  {author} {\bibinfo {author} {\bibfnamefont {D.}~\bibnamefont
  {Paloli}}, \bibinfo {author} {\bibfnamefont {P.~S.}\ \bibnamefont {Mohanty}},
  \bibinfo {author} {\bibfnamefont {J.~J.}\ \bibnamefont {Crassous}}, \bibinfo
  {author} {\bibfnamefont {E.}~\bibnamefont {Zaccarelli}}, \ and\ \bibinfo
  {author} {\bibfnamefont {P.}~\bibnamefont {Schurtenberger}},\ }\href@noop {}
  {\bibfield  {journal} {\bibinfo  {journal} {Soft Matter}\ }\textbf {\bibinfo
  {volume} {9}},\ \bibinfo {pages} {3000} (\bibinfo {year} {2013})}\BibitemShut
  {NoStop}%
\bibitem [{\citenamefont {Del~Monte}\ and\ \citenamefont
  {Zaccarelli}(2024)}]{del2024numerical}%
  \BibitemOpen
  \bibfield  {author} {\bibinfo {author} {\bibfnamefont {G.}~\bibnamefont
  {Del~Monte}}\ and\ \bibinfo {author} {\bibfnamefont {E.}~\bibnamefont
  {Zaccarelli}},\ }\href@noop {} {\bibfield  {journal} {\bibinfo  {journal}
  {Phys. Rev. X}\ }\textbf {\bibinfo {volume} {14}},\ \bibinfo {pages} {041067}
  (\bibinfo {year} {2024})}\BibitemShut {NoStop}%
\bibitem [{\citenamefont {Bergman}\ \emph {et~al.}(2018)\citenamefont
  {Bergman}, \citenamefont {Gnan}, \citenamefont {Obiols-Rabasa}, \citenamefont
  {Meijer}, \citenamefont {Rovigatti}, \citenamefont {Zaccarelli},\ and\
  \citenamefont {Schurtenberger}}]{bergman2018new}%
  \BibitemOpen
  \bibfield  {author} {\bibinfo {author} {\bibfnamefont {M.~J.}\ \bibnamefont
  {Bergman}}, \bibinfo {author} {\bibfnamefont {N.}~\bibnamefont {Gnan}},
  \bibinfo {author} {\bibfnamefont {M.}~\bibnamefont {Obiols-Rabasa}}, \bibinfo
  {author} {\bibfnamefont {J.-M.}\ \bibnamefont {Meijer}}, \bibinfo {author}
  {\bibfnamefont {L.}~\bibnamefont {Rovigatti}}, \bibinfo {author}
  {\bibfnamefont {E.}~\bibnamefont {Zaccarelli}}, \ and\ \bibinfo {author}
  {\bibfnamefont {P.}~\bibnamefont {Schurtenberger}},\ }\href@noop {}
  {\bibfield  {journal} {\bibinfo  {journal} {Nature Communications}\ }\textbf
  {\bibinfo {volume} {9}},\ \bibinfo {pages} {5039} (\bibinfo {year}
  {2018})}\BibitemShut {NoStop}%
\bibitem [{\citenamefont {Stieger}\ \emph {et~al.}(2004)\citenamefont
  {Stieger}, \citenamefont {Richtering}, \citenamefont {Pedersen},\ and\
  \citenamefont {Lindner}}]{Sti04}%
  \BibitemOpen
  \bibfield  {author} {\bibinfo {author} {\bibfnamefont {M.}~\bibnamefont
  {Stieger}}, \bibinfo {author} {\bibfnamefont {W.}~\bibnamefont {Richtering}},
  \bibinfo {author} {\bibfnamefont {J.~S.}\ \bibnamefont {Pedersen}}, \ and\
  \bibinfo {author} {\bibfnamefont {P.}~\bibnamefont {Lindner}},\ }\href@noop
  {} {\bibfield  {journal} {\bibinfo  {journal} {The Journal of chemical
  physics}\ }\textbf {\bibinfo {volume} {120}},\ \bibinfo {pages} {6197}
  (\bibinfo {year} {2004})}\BibitemShut {NoStop}%
\bibitem [{\citenamefont {Hazra}\ \emph {et~al.}(2023)\citenamefont {Hazra},
  \citenamefont {Ninarello}, \citenamefont {Scotti}, \citenamefont {Houston},
  \citenamefont {Mota-Santiago}, \citenamefont {Zaccarelli},\ and\
  \citenamefont {Crassous}}]{Haz23}%
  \BibitemOpen
  \bibfield  {author} {\bibinfo {author} {\bibfnamefont {N.}~\bibnamefont
  {Hazra}}, \bibinfo {author} {\bibfnamefont {A.}~\bibnamefont {Ninarello}},
  \bibinfo {author} {\bibfnamefont {A.}~\bibnamefont {Scotti}}, \bibinfo
  {author} {\bibfnamefont {J.~E.}\ \bibnamefont {Houston}}, \bibinfo {author}
  {\bibfnamefont {P.}~\bibnamefont {Mota-Santiago}}, \bibinfo {author}
  {\bibfnamefont {E.}~\bibnamefont {Zaccarelli}}, \ and\ \bibinfo {author}
  {\bibfnamefont {J.~J.}\ \bibnamefont {Crassous}},\ }\href@noop {} {\bibfield
  {journal} {\bibinfo  {journal} {Macromolecules}\ }\textbf {\bibinfo {volume}
  {57}},\ \bibinfo {pages} {339} (\bibinfo {year} {2023})}\BibitemShut
  {NoStop}%
\bibitem [{\citenamefont {Camerin}\ \emph {et~al.}(2020)\citenamefont
  {Camerin}, \citenamefont {Gnan}, \citenamefont {Ruiz-Franco}, \citenamefont
  {Ninarello}, \citenamefont {Rovigatti},\ and\ \citenamefont
  {Zaccarelli}}]{Cam20}%
  \BibitemOpen
  \bibfield  {author} {\bibinfo {author} {\bibfnamefont {F.}~\bibnamefont
  {Camerin}}, \bibinfo {author} {\bibfnamefont {N.}~\bibnamefont {Gnan}},
  \bibinfo {author} {\bibfnamefont {J.}~\bibnamefont {Ruiz-Franco}}, \bibinfo
  {author} {\bibfnamefont {A.}~\bibnamefont {Ninarello}}, \bibinfo {author}
  {\bibfnamefont {L.}~\bibnamefont {Rovigatti}}, \ and\ \bibinfo {author}
  {\bibfnamefont {E.}~\bibnamefont {Zaccarelli}},\ }\href@noop {} {\bibfield
  {journal} {\bibinfo  {journal} {Physical Review X}\ }\textbf {\bibinfo
  {volume} {10}},\ \bibinfo {pages} {031012} (\bibinfo {year}
  {2020})}\BibitemShut {NoStop}%
\bibitem [{\citenamefont {Berthier}\ \emph {et~al.}(2010)\citenamefont
  {Berthier}, \citenamefont {Moreno},\ and\ \citenamefont {Szamel}}]{Ber10}%
  \BibitemOpen
  \bibfield  {author} {\bibinfo {author} {\bibfnamefont {L.}~\bibnamefont
  {Berthier}}, \bibinfo {author} {\bibfnamefont {A.~J.}\ \bibnamefont
  {Moreno}}, \ and\ \bibinfo {author} {\bibfnamefont {G.}~\bibnamefont
  {Szamel}},\ }\href@noop {} {\bibfield  {journal} {\bibinfo  {journal}
  {Physical Review E}\ }\textbf {\bibinfo {volume} {82}},\ \bibinfo {pages}
  {060501} (\bibinfo {year} {2010})}\BibitemShut {NoStop}%
\bibitem [{\citenamefont {Wood}\ \emph {et~al.}(2023)\citenamefont {Wood},
  \citenamefont {Razinkov}, \citenamefont {Qi}, \citenamefont {Roberts},
  \citenamefont {Vermant},\ and\ \citenamefont {Furst}}]{Woo23}%
  \BibitemOpen
  \bibfield  {author} {\bibinfo {author} {\bibfnamefont {C.~V.}\ \bibnamefont
  {Wood}}, \bibinfo {author} {\bibfnamefont {V.~I.}\ \bibnamefont {Razinkov}},
  \bibinfo {author} {\bibfnamefont {W.}~\bibnamefont {Qi}}, \bibinfo {author}
  {\bibfnamefont {C.~J.}\ \bibnamefont {Roberts}}, \bibinfo {author}
  {\bibfnamefont {J.}~\bibnamefont {Vermant}}, \ and\ \bibinfo {author}
  {\bibfnamefont {E.~M.}\ \bibnamefont {Furst}},\ }\href@noop {} {\bibfield
  {journal} {\bibinfo  {journal} {Langmuir}\ }\textbf {\bibinfo {volume}
  {39}},\ \bibinfo {pages} {7775} (\bibinfo {year} {2023})}\BibitemShut
  {NoStop}%
\bibitem [{\citenamefont {Freer}\ \emph {et~al.}(2004)\citenamefont {Freer},
  \citenamefont {Yim}, \citenamefont {Fuller},\ and\ \citenamefont
  {Radke}}]{Fre04}%
  \BibitemOpen
  \bibfield  {author} {\bibinfo {author} {\bibfnamefont {E.~M.}\ \bibnamefont
  {Freer}}, \bibinfo {author} {\bibfnamefont {K.~S.}\ \bibnamefont {Yim}},
  \bibinfo {author} {\bibfnamefont {G.~G.}\ \bibnamefont {Fuller}}, \ and\
  \bibinfo {author} {\bibfnamefont {C.~J.}\ \bibnamefont {Radke}},\ }\href@noop
  {} {\bibfield  {journal} {\bibinfo  {journal} {The Journal of Physical
  Chemistry B}\ }\textbf {\bibinfo {volume} {108}},\ \bibinfo {pages} {3835}
  (\bibinfo {year} {2004})}\BibitemShut {NoStop}%
\bibitem [{\citenamefont {Bos}\ and\ \citenamefont {Van~Vliet}(2001)}]{Bos01}%
  \BibitemOpen
  \bibfield  {author} {\bibinfo {author} {\bibfnamefont {M.~A.}\ \bibnamefont
  {Bos}}\ and\ \bibinfo {author} {\bibfnamefont {T.}~\bibnamefont
  {Van~Vliet}},\ }\href@noop {} {\bibfield  {journal} {\bibinfo  {journal}
  {Advances in colloid and interface science}\ }\textbf {\bibinfo {volume}
  {91}},\ \bibinfo {pages} {437} (\bibinfo {year} {2001})}\BibitemShut
  {NoStop}%
\bibitem [{\citenamefont {Bochenek}\ \emph {et~al.}(2023)\citenamefont
  {Bochenek}, \citenamefont {Rudov}, \citenamefont {Sassmann}, \citenamefont
  {Potemkin},\ and\ \citenamefont {Richtering}}]{Boc23}%
  \BibitemOpen
  \bibfield  {author} {\bibinfo {author} {\bibfnamefont {S.}~\bibnamefont
  {Bochenek}}, \bibinfo {author} {\bibfnamefont {A.~A.}\ \bibnamefont {Rudov}},
  \bibinfo {author} {\bibfnamefont {T.}~\bibnamefont {Sassmann}}, \bibinfo
  {author} {\bibfnamefont {I.~I.}\ \bibnamefont {Potemkin}}, \ and\ \bibinfo
  {author} {\bibfnamefont {W.}~\bibnamefont {Richtering}},\ }\href@noop {}
  {\bibfield  {journal} {\bibinfo  {journal} {Langmuir}\ }\textbf {\bibinfo
  {volume} {39}},\ \bibinfo {pages} {18354} (\bibinfo {year}
  {2023})}\BibitemShut {NoStop}%
\bibitem [{\citenamefont {Kuk}\ \emph {et~al.}(2023)\citenamefont {Kuk},
  \citenamefont {Abgarjan}, \citenamefont {Gregel}, \citenamefont {Zhou},
  \citenamefont {Fadanelli}, \citenamefont {Buttinoni},\ and\ \citenamefont
  {Karg}}]{Kuk23}%
  \BibitemOpen
  \bibfield  {author} {\bibinfo {author} {\bibfnamefont {K.}~\bibnamefont
  {Kuk}}, \bibinfo {author} {\bibfnamefont {V.}~\bibnamefont {Abgarjan}},
  \bibinfo {author} {\bibfnamefont {L.}~\bibnamefont {Gregel}}, \bibinfo
  {author} {\bibfnamefont {Y.}~\bibnamefont {Zhou}}, \bibinfo {author}
  {\bibfnamefont {V.~C.}\ \bibnamefont {Fadanelli}}, \bibinfo {author}
  {\bibfnamefont {I.}~\bibnamefont {Buttinoni}}, \ and\ \bibinfo {author}
  {\bibfnamefont {M.}~\bibnamefont {Karg}},\ }\href@noop {} {\bibfield
  {journal} {\bibinfo  {journal} {Soft Matter}\ }\textbf {\bibinfo {volume}
  {19}},\ \bibinfo {pages} {175} (\bibinfo {year} {2023})}\BibitemShut
  {NoStop}%
\bibitem [{\citenamefont {Rey}\ \emph {et~al.}(2016)\citenamefont {Rey},
  \citenamefont {Fern{\'a}ndez-Rodr{\'\i}guez}, \citenamefont {Steinacher},
  \citenamefont {Scheidegger}, \citenamefont {Geisel}, \citenamefont
  {Richtering}, \citenamefont {Squires},\ and\ \citenamefont {Isa}}]{Rey16}%
  \BibitemOpen
  \bibfield  {author} {\bibinfo {author} {\bibfnamefont {M.}~\bibnamefont
  {Rey}}, \bibinfo {author} {\bibfnamefont {M.~{\'A}.}\ \bibnamefont
  {Fern{\'a}ndez-Rodr{\'\i}guez}}, \bibinfo {author} {\bibfnamefont
  {M.}~\bibnamefont {Steinacher}}, \bibinfo {author} {\bibfnamefont
  {L.}~\bibnamefont {Scheidegger}}, \bibinfo {author} {\bibfnamefont
  {K.}~\bibnamefont {Geisel}}, \bibinfo {author} {\bibfnamefont
  {W.}~\bibnamefont {Richtering}}, \bibinfo {author} {\bibfnamefont {T.~M.}\
  \bibnamefont {Squires}}, \ and\ \bibinfo {author} {\bibfnamefont
  {L.}~\bibnamefont {Isa}},\ }\href@noop {} {\bibfield  {journal} {\bibinfo
  {journal} {Soft Matter}\ }\textbf {\bibinfo {volume} {12}},\ \bibinfo {pages}
  {3545} (\bibinfo {year} {2016})}\BibitemShut {NoStop}%
\bibitem [{\citenamefont {Geisel}\ \emph {et~al.}(2014)\citenamefont {Geisel},
  \citenamefont {Isa},\ and\ \citenamefont {Richtering}}]{Gei14}%
  \BibitemOpen
  \bibfield  {author} {\bibinfo {author} {\bibfnamefont {K.}~\bibnamefont
  {Geisel}}, \bibinfo {author} {\bibfnamefont {L.}~\bibnamefont {Isa}}, \ and\
  \bibinfo {author} {\bibfnamefont {W.}~\bibnamefont {Richtering}},\
  }\href@noop {} {\bibfield  {journal} {\bibinfo  {journal} {Angewandte
  Chemie}\ }\textbf {\bibinfo {volume} {126}},\ \bibinfo {pages} {5005}
  (\bibinfo {year} {2014})}\BibitemShut {NoStop}%
\bibitem [{\citenamefont {Bochenek}\ \emph {et~al.}(2021)\citenamefont
  {Bochenek}, \citenamefont {Scotti},\ and\ \citenamefont
  {Richtering}}]{Boc21}%
  \BibitemOpen
  \bibfield  {author} {\bibinfo {author} {\bibfnamefont {S.}~\bibnamefont
  {Bochenek}}, \bibinfo {author} {\bibfnamefont {A.}~\bibnamefont {Scotti}}, \
  and\ \bibinfo {author} {\bibfnamefont {W.}~\bibnamefont {Richtering}},\
  }\href@noop {} {\bibfield  {journal} {\bibinfo  {journal} {Soft Matter}\
  }\textbf {\bibinfo {volume} {17}},\ \bibinfo {pages} {976} (\bibinfo {year}
  {2021})}\BibitemShut {NoStop}%
\bibitem [{\citenamefont {P{\`a}mies}\ \emph {et~al.}(2009)\citenamefont
  {P{\`a}mies}, \citenamefont {Cacciuto},\ and\ \citenamefont
  {Frenkel}}]{Pam09}%
  \BibitemOpen
  \bibfield  {author} {\bibinfo {author} {\bibfnamefont {J.~C.}\ \bibnamefont
  {P{\`a}mies}}, \bibinfo {author} {\bibfnamefont {A.}~\bibnamefont
  {Cacciuto}}, \ and\ \bibinfo {author} {\bibfnamefont {D.}~\bibnamefont
  {Frenkel}},\ }\href@noop {} {\bibfield  {journal} {\bibinfo  {journal} {The
  Journal of chemical physics}\ }\textbf {\bibinfo {volume} {131}} (\bibinfo
  {year} {2009})}\BibitemShut {NoStop}%
\bibitem [{\citenamefont {Pellet}\ and\ \citenamefont {Cloitre}(2016)}]{Pel16}%
  \BibitemOpen
  \bibfield  {author} {\bibinfo {author} {\bibfnamefont {C.}~\bibnamefont
  {Pellet}}\ and\ \bibinfo {author} {\bibfnamefont {M.}~\bibnamefont
  {Cloitre}},\ }\href@noop {} {\bibfield  {journal} {\bibinfo  {journal} {Soft
  Matter}\ }\textbf {\bibinfo {volume} {12}},\ \bibinfo {pages} {3710}
  (\bibinfo {year} {2016})}\BibitemShut {NoStop}%
\bibitem [{\citenamefont {Tatry}\ \emph {et~al.}(2023)\citenamefont {Tatry},
  \citenamefont {Laurichesse}, \citenamefont {Vermant}, \citenamefont
  {Ravaine},\ and\ \citenamefont {Schmitt}}]{Tat23}%
  \BibitemOpen
  \bibfield  {author} {\bibinfo {author} {\bibfnamefont {M.-C.}\ \bibnamefont
  {Tatry}}, \bibinfo {author} {\bibfnamefont {E.}~\bibnamefont {Laurichesse}},
  \bibinfo {author} {\bibfnamefont {J.}~\bibnamefont {Vermant}}, \bibinfo
  {author} {\bibfnamefont {V.}~\bibnamefont {Ravaine}}, \ and\ \bibinfo
  {author} {\bibfnamefont {V.}~\bibnamefont {Schmitt}},\ }\href@noop {}
  {\bibfield  {journal} {\bibinfo  {journal} {Journal of Colloid and Interface
  Science}\ }\textbf {\bibinfo {volume} {629}},\ \bibinfo {pages} {288}
  (\bibinfo {year} {2023})}\BibitemShut {NoStop}%
\bibitem [{\citenamefont {Conley}\ \emph {et~al.}(2019)\citenamefont {Conley},
  \citenamefont {Zhang}, \citenamefont {Aebischer}, \citenamefont {Harden},\
  and\ \citenamefont {Scheffold}}]{Con19}%
  \BibitemOpen
  \bibfield  {author} {\bibinfo {author} {\bibfnamefont {G.~M.}\ \bibnamefont
  {Conley}}, \bibinfo {author} {\bibfnamefont {C.}~\bibnamefont {Zhang}},
  \bibinfo {author} {\bibfnamefont {P.}~\bibnamefont {Aebischer}}, \bibinfo
  {author} {\bibfnamefont {J.~L.}\ \bibnamefont {Harden}}, \ and\ \bibinfo
  {author} {\bibfnamefont {F.}~\bibnamefont {Scheffold}},\ }\href@noop {}
  {\bibfield  {journal} {\bibinfo  {journal} {Nature Communications}\ }\textbf
  {\bibinfo {volume} {10}},\ \bibinfo {pages} {2436} (\bibinfo {year}
  {2019})}\BibitemShut {NoStop}%
\bibitem [{\citenamefont {Bergman}\ \emph {et~al.}(2025)\citenamefont
  {Bergman}, \citenamefont {Xu}, \citenamefont {Muñéton~Díaz}, \citenamefont
  {Zhang}, \citenamefont {Mason},\ and\ \citenamefont
  {Scheffold}}]{bergman2025free}%
  \BibitemOpen
  \bibfield  {author} {\bibinfo {author} {\bibfnamefont {M.}~\bibnamefont
  {Bergman}}, \bibinfo {author} {\bibfnamefont {Y.}~\bibnamefont {Xu}},
  \bibinfo {author} {\bibfnamefont {J.}~\bibnamefont {Muñéton~Díaz}},
  \bibinfo {author} {\bibfnamefont {C.}~\bibnamefont {Zhang}}, \bibinfo
  {author} {\bibfnamefont {T.~G.}\ \bibnamefont {Mason}}, \ and\ \bibinfo
  {author} {\bibfnamefont {F.}~\bibnamefont {Scheffold}},\ }\href@noop {}
  {\bibfield  {journal} {\bibinfo  {journal} {Langmuir}\ } (\bibinfo {year}
  {2025})}\BibitemShut {NoStop}%
\bibitem [{\citenamefont {Zaccarelli}\ \emph {et~al.}(2001)\citenamefont
  {Zaccarelli}, \citenamefont {Foffi}, \citenamefont {Dawson}, \citenamefont
  {Sciortino},\ and\ \citenamefont {Tartaglia}}]{Zac01}%
  \BibitemOpen
  \bibfield  {author} {\bibinfo {author} {\bibfnamefont {E.}~\bibnamefont
  {Zaccarelli}}, \bibinfo {author} {\bibfnamefont {G.}~\bibnamefont {Foffi}},
  \bibinfo {author} {\bibfnamefont {K.}~\bibnamefont {Dawson}}, \bibinfo
  {author} {\bibfnamefont {F.}~\bibnamefont {Sciortino}}, \ and\ \bibinfo
  {author} {\bibfnamefont {P.}~\bibnamefont {Tartaglia}},\ }\href@noop {}
  {\bibfield  {journal} {\bibinfo  {journal} {Physical Review E}\ }\textbf
  {\bibinfo {volume} {63}},\ \bibinfo {pages} {031501} (\bibinfo {year}
  {2001})}\BibitemShut {NoStop}%
\bibitem [{\citenamefont {Pham}\ \emph {et~al.}(2008)\citenamefont {Pham},
  \citenamefont {Petekidis}, \citenamefont {Vlassopoulos}, \citenamefont
  {Egelhaaf}, \citenamefont {Poon},\ and\ \citenamefont {Pusey}}]{Pha08}%
  \BibitemOpen
  \bibfield  {author} {\bibinfo {author} {\bibfnamefont {K.}~\bibnamefont
  {Pham}}, \bibinfo {author} {\bibfnamefont {G.}~\bibnamefont {Petekidis}},
  \bibinfo {author} {\bibfnamefont {D.}~\bibnamefont {Vlassopoulos}}, \bibinfo
  {author} {\bibfnamefont {S.}~\bibnamefont {Egelhaaf}}, \bibinfo {author}
  {\bibfnamefont {W.}~\bibnamefont {Poon}}, \ and\ \bibinfo {author}
  {\bibfnamefont {P.}~\bibnamefont {Pusey}},\ }\href@noop {} {\bibfield
  {journal} {\bibinfo  {journal} {Journal of Rheology}\ }\textbf {\bibinfo
  {volume} {52}},\ \bibinfo {pages} {649} (\bibinfo {year} {2008})}\BibitemShut
  {NoStop}%
\bibitem [{\citenamefont {Zou}\ \emph {et~al.}(2024)\citenamefont {Zou},
  \citenamefont {Ruan}, \citenamefont {Zhang},\ and\ \citenamefont
  {Liu}}]{zou2024tunable}%
  \BibitemOpen
  \bibfield  {author} {\bibinfo {author} {\bibfnamefont {Q.}~\bibnamefont
  {Zou}}, \bibinfo {author} {\bibfnamefont {Y.}~\bibnamefont {Ruan}}, \bibinfo
  {author} {\bibfnamefont {R.}~\bibnamefont {Zhang}}, \ and\ \bibinfo {author}
  {\bibfnamefont {G.}~\bibnamefont {Liu}},\ }\href@noop {} {\bibfield
  {journal} {\bibinfo  {journal} {Macromolecules}\ }\textbf {\bibinfo {volume}
  {57}},\ \bibinfo {pages} {777} (\bibinfo {year} {2024})}\BibitemShut
  {NoStop}%
\bibitem [{\citenamefont {Wu}\ \emph {et~al.}(2018)\citenamefont {Wu},
  \citenamefont {Aroush}, \citenamefont {Asnacios}, \citenamefont {Chen},
  \citenamefont {Dokukin}, \citenamefont {Doss}, \citenamefont {Durand-Smet},
  \citenamefont {Ekpenyong}, \citenamefont {Guck}, \citenamefont {Guz} \emph
  {et~al.}}]{wu2018comparison}%
  \BibitemOpen
  \bibfield  {author} {\bibinfo {author} {\bibfnamefont {P.-H.}\ \bibnamefont
  {Wu}}, \bibinfo {author} {\bibfnamefont {D.~R.-B.}\ \bibnamefont {Aroush}},
  \bibinfo {author} {\bibfnamefont {A.}~\bibnamefont {Asnacios}}, \bibinfo
  {author} {\bibfnamefont {W.-C.}\ \bibnamefont {Chen}}, \bibinfo {author}
  {\bibfnamefont {M.~E.}\ \bibnamefont {Dokukin}}, \bibinfo {author}
  {\bibfnamefont {B.~L.}\ \bibnamefont {Doss}}, \bibinfo {author}
  {\bibfnamefont {P.}~\bibnamefont {Durand-Smet}}, \bibinfo {author}
  {\bibfnamefont {A.}~\bibnamefont {Ekpenyong}}, \bibinfo {author}
  {\bibfnamefont {J.}~\bibnamefont {Guck}}, \bibinfo {author} {\bibfnamefont
  {N.~V.}\ \bibnamefont {Guz}},  \emph {et~al.},\ }\href@noop {} {\bibfield
  {journal} {\bibinfo  {journal} {Nature methods}\ }\textbf {\bibinfo {volume}
  {15}},\ \bibinfo {pages} {491} (\bibinfo {year} {2018})}\BibitemShut
  {NoStop}%
\bibitem [{\citenamefont {Komaragiri}\ \emph {et~al.}(2024)\citenamefont
  {Komaragiri}, \citenamefont {Pires}, \citenamefont {Spiegler}, \citenamefont
  {Dau}, \citenamefont {Biedenweg}, \citenamefont {Salas}, \citenamefont
  {Hossain}, \citenamefont {Fregin}, \citenamefont {Gross}, \citenamefont
  {Gellert} \emph {et~al.}}]{komaragiri2024cell}%
  \BibitemOpen
  \bibfield  {author} {\bibinfo {author} {\bibfnamefont {Y.}~\bibnamefont
  {Komaragiri}}, \bibinfo {author} {\bibfnamefont {R.~H.}\ \bibnamefont
  {Pires}}, \bibinfo {author} {\bibfnamefont {S.}~\bibnamefont {Spiegler}},
  \bibinfo {author} {\bibfnamefont {H.~T.}\ \bibnamefont {Dau}}, \bibinfo
  {author} {\bibfnamefont {D.}~\bibnamefont {Biedenweg}}, \bibinfo {author}
  {\bibfnamefont {C.~O.}\ \bibnamefont {Salas}}, \bibinfo {author}
  {\bibfnamefont {M.~F.}\ \bibnamefont {Hossain}}, \bibinfo {author}
  {\bibfnamefont {B.}~\bibnamefont {Fregin}}, \bibinfo {author} {\bibfnamefont
  {S.}~\bibnamefont {Gross}}, \bibinfo {author} {\bibfnamefont
  {M.}~\bibnamefont {Gellert}},  \emph {et~al.},\ }\href@noop {} {\bibfield
  {journal} {\bibinfo  {journal} {Communications Physics}\ }\textbf {\bibinfo
  {volume} {7}},\ \bibinfo {pages} {252} (\bibinfo {year} {2024})}\BibitemShut
  {NoStop}%
\bibitem [{\citenamefont {Scotti}\ \emph
  {et~al.}(2019{\natexlab{b}})\citenamefont {Scotti}, \citenamefont {Denton},
  \citenamefont {Brugnoni}, \citenamefont {Houston}, \citenamefont {Schweins},
  \citenamefont {Potemkin},\ and\ \citenamefont {Richtering}}]{Sco19b}%
  \BibitemOpen
  \bibfield  {author} {\bibinfo {author} {\bibfnamefont {A.}~\bibnamefont
  {Scotti}}, \bibinfo {author} {\bibfnamefont {A.~R.}\ \bibnamefont {Denton}},
  \bibinfo {author} {\bibfnamefont {M.}~\bibnamefont {Brugnoni}}, \bibinfo
  {author} {\bibfnamefont {J.~E.}\ \bibnamefont {Houston}}, \bibinfo {author}
  {\bibfnamefont {R.}~\bibnamefont {Schweins}}, \bibinfo {author}
  {\bibfnamefont {I.~I.}\ \bibnamefont {Potemkin}}, \ and\ \bibinfo {author}
  {\bibfnamefont {W.}~\bibnamefont {Richtering}},\ }\href@noop {} {\bibfield
  {journal} {\bibinfo  {journal} {Macromolecules}\ }\textbf {\bibinfo {volume}
  {52}},\ \bibinfo {pages} {3995} (\bibinfo {year}
  {2019}{\natexlab{b}})}\BibitemShut {NoStop}%
\bibitem [{\citenamefont {Pelton}\ and\ \citenamefont
  {Chibante}(1986)}]{Pel86}%
  \BibitemOpen
  \bibfield  {author} {\bibinfo {author} {\bibfnamefont {R.}~\bibnamefont
  {Pelton}}\ and\ \bibinfo {author} {\bibfnamefont {P.}~\bibnamefont
  {Chibante}},\ }\href@noop {} {\bibfield  {journal} {\bibinfo  {journal}
  {Colloids and Surfaces}\ }\textbf {\bibinfo {volume} {20}},\ \bibinfo {pages}
  {247} (\bibinfo {year} {1986})}\BibitemShut {NoStop}%
\bibitem [{\citenamefont {Vandebril}\ \emph {et~al.}(2010)\citenamefont
  {Vandebril}, \citenamefont {Franck}, \citenamefont {Fuller}, \citenamefont
  {Moldenaers},\ and\ \citenamefont {Vermant}}]{Van10}%
  \BibitemOpen
  \bibfield  {author} {\bibinfo {author} {\bibfnamefont {S.}~\bibnamefont
  {Vandebril}}, \bibinfo {author} {\bibfnamefont {A.}~\bibnamefont {Franck}},
  \bibinfo {author} {\bibfnamefont {G.~G.}\ \bibnamefont {Fuller}}, \bibinfo
  {author} {\bibfnamefont {P.}~\bibnamefont {Moldenaers}}, \ and\ \bibinfo
  {author} {\bibfnamefont {J.}~\bibnamefont {Vermant}},\ }\href@noop {}
  {\bibfield  {journal} {\bibinfo  {journal} {Rheologica Acta}\ }\textbf
  {\bibinfo {volume} {49}},\ \bibinfo {pages} {131} (\bibinfo {year}
  {2010})}\BibitemShut {NoStop}%
\bibitem [{\citenamefont {Thompson}\ \emph {et~al.}(2022)\citenamefont
  {Thompson}, \citenamefont {Aktulga}, \citenamefont {Berger}, \citenamefont
  {Bolintineanu}, \citenamefont {Brown}, \citenamefont {Crozier}, \citenamefont
  {in~'t Veld}, \citenamefont {Kohlmeyer}, \citenamefont {Moore}, \citenamefont
  {Nguyen}, \citenamefont {Shan}, \citenamefont {Stevens}, \citenamefont
  {Tranchida}, \citenamefont {Trott},\ and\ \citenamefont {Plimpton}}]{LAMMPS}%
  \BibitemOpen
  \bibfield  {author} {\bibinfo {author} {\bibfnamefont {A.~P.}\ \bibnamefont
  {Thompson}}, \bibinfo {author} {\bibfnamefont {H.~M.}\ \bibnamefont
  {Aktulga}}, \bibinfo {author} {\bibfnamefont {R.}~\bibnamefont {Berger}},
  \bibinfo {author} {\bibfnamefont {D.~S.}\ \bibnamefont {Bolintineanu}},
  \bibinfo {author} {\bibfnamefont {W.~M.}\ \bibnamefont {Brown}}, \bibinfo
  {author} {\bibfnamefont {P.~S.}\ \bibnamefont {Crozier}}, \bibinfo {author}
  {\bibfnamefont {P.~J.}\ \bibnamefont {in~'t Veld}}, \bibinfo {author}
  {\bibfnamefont {A.}~\bibnamefont {Kohlmeyer}}, \bibinfo {author}
  {\bibfnamefont {S.~G.}\ \bibnamefont {Moore}}, \bibinfo {author}
  {\bibfnamefont {T.~D.}\ \bibnamefont {Nguyen}}, \bibinfo {author}
  {\bibfnamefont {R.}~\bibnamefont {Shan}}, \bibinfo {author} {\bibfnamefont
  {M.~J.}\ \bibnamefont {Stevens}}, \bibinfo {author} {\bibfnamefont
  {J.}~\bibnamefont {Tranchida}}, \bibinfo {author} {\bibfnamefont
  {C.}~\bibnamefont {Trott}}, \ and\ \bibinfo {author} {\bibfnamefont {S.~J.}\
  \bibnamefont {Plimpton}},\ }\href {\doibase 10.1016/j.cpc.2021.108171}
  {\bibfield  {journal} {\bibinfo  {journal} {Comp. Phys. Comm.}\ }\textbf
  {\bibinfo {volume} {271}},\ \bibinfo {pages} {108171} (\bibinfo {year}
  {2022})}\BibitemShut {NoStop}%
\end{thebibliography}%

\noindent\textbf{Acknowledgements}\\
\noindent J.R.-F and T.H contributed equally to this work. A.S acknowledges financial support from the Knut and Alice Wallenberg Foundation (Wallenberg Academy Fellows). EZ acknowledges financial support from EU MSCA Doctoral Network QLUSTER, Grant Agreement 101072964. The authors gratefully acknowledge financial support from the Deutsche Forschungsgemeinschaft (DFG) withing the SFB 985 ``Functional microgels and microgel systems''. The authors  thank the RWTH Aachen University that granted the computation time under project rwth0946. J.R.-F thanks the Department of Research and Universities of the Generalitat de Catalunya through Beatriu de Pin\'os program Grant (Contract No. 2022 BP 00156).\\

\onecolumngrid

\clearpage

\section*{Supplemental Material for \\ ```Soft ULC Microgels at the Interface Interact and Flow as Hertzian-like Colloids''}
\setcounter{equation}{0}
\setcounter{figure}{0}
\setcounter{table}{0}
\setcounter{section}{0}

\renewcommand\thefigure{S\arabic{figure}} 
\noindent

\section{Flow curves}

\begin{figure}[htbp!]
    \centering
    \includegraphics[width=\textwidth]{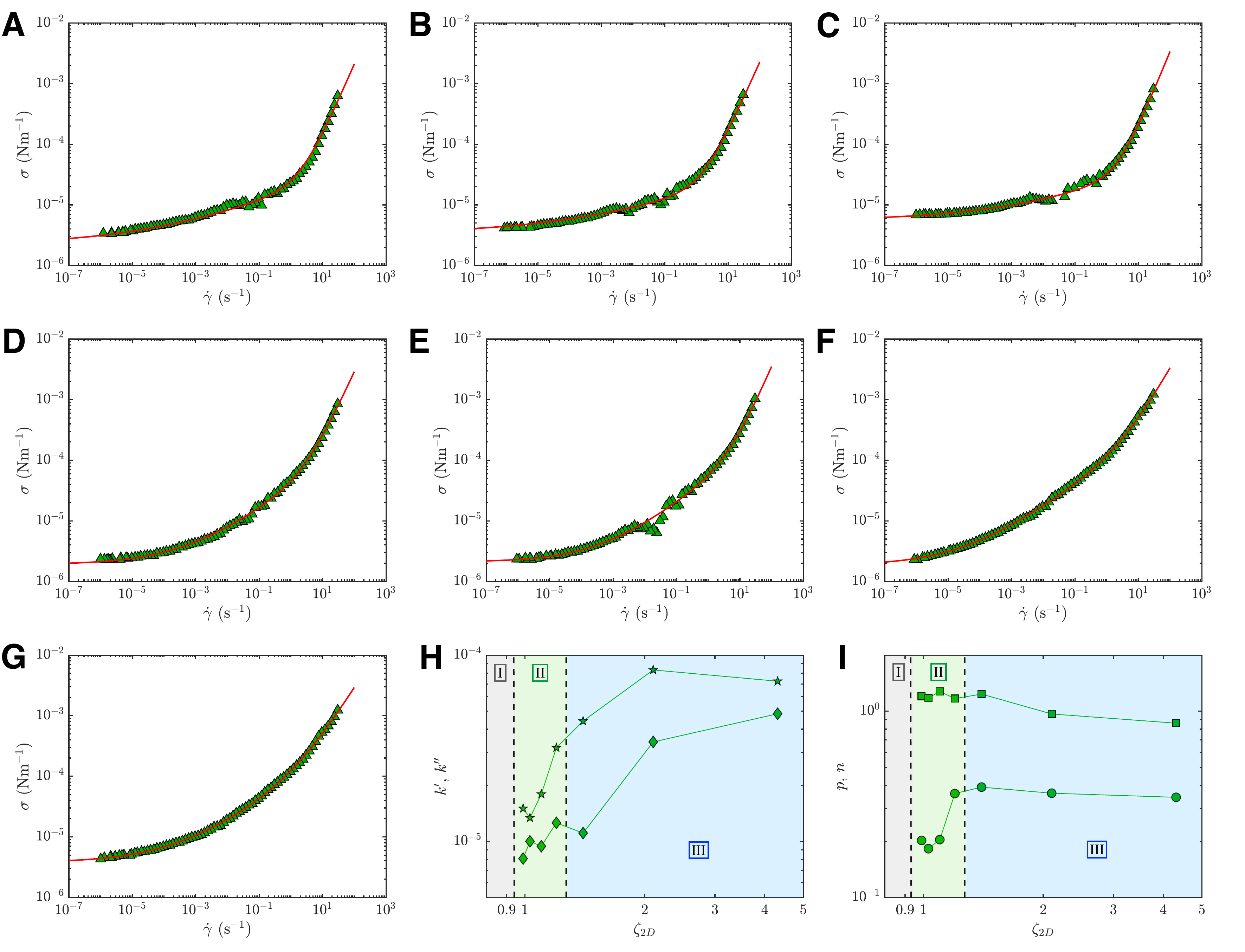}
    \caption{\textbf{Flow curves of ULC microgel monolayers at the oil-water interface.} (\textbf{A-F}) Values of the shear stress (squares), $\sigma$, \emph{vs.}~the shear-rate, $\dot\gamma$, for the soft microgels at different values of $\zeta_{2D}$: (\textbf{A}) $0.99\pm0.21$, (\textbf{B}) $1.03\pm0.22$, (\textbf{C}) $1.10\pm0.23$, (\textbf{D}) $1.20\pm0.25$, (\textbf{E}) $1.40\pm0.30$, (\textbf{F}) $2.10\pm0.44$, (\textbf{G}) $4.31\pm0.92$. The solid lines represent fit with the Herschel–Bulkley model, Eq.~2. Parameters obtained from fitting the experimental flow curves with $\sigma/\sigma_y = 1 + k\tilde{\gamma}^u + k'\tilde{\gamma}^p$: (\textbf{H}) prefactors $k$ (circles) and $k^\prime$ (squares) and (i) exponents $u$ (squares) and $p$ (circles) as a function of $\zeta_{2D}$. Vertical dashed lines mark the different regions of the compression isotherms identified in Fig.~1(\textbf{O}) of the manuscript.}
    \label{fig:ULC3FC_SI}
\end{figure}

Figs.~\ref{fig:ULC3FC_SI}(a)-to-\ref{fig:ULC3FC_SI}(g) show the flow curves for the monolayer at different compressions, i.e.~different $\zeta_{2D}$, and the relative fits with Eq. 2 in the main text (solid lines).
Figs.~\ref{fig:ULC3FC_SI}(h) and~\ref{fig:ULC3FC_SI}(i) show the evolution of the fitting parameter of Eq. 2.

\newpage

\section{Amplitude sweeps}\label{sec:AS}

\begin{figure}[htbp!]
    \centering
    \includegraphics[width=\textwidth]{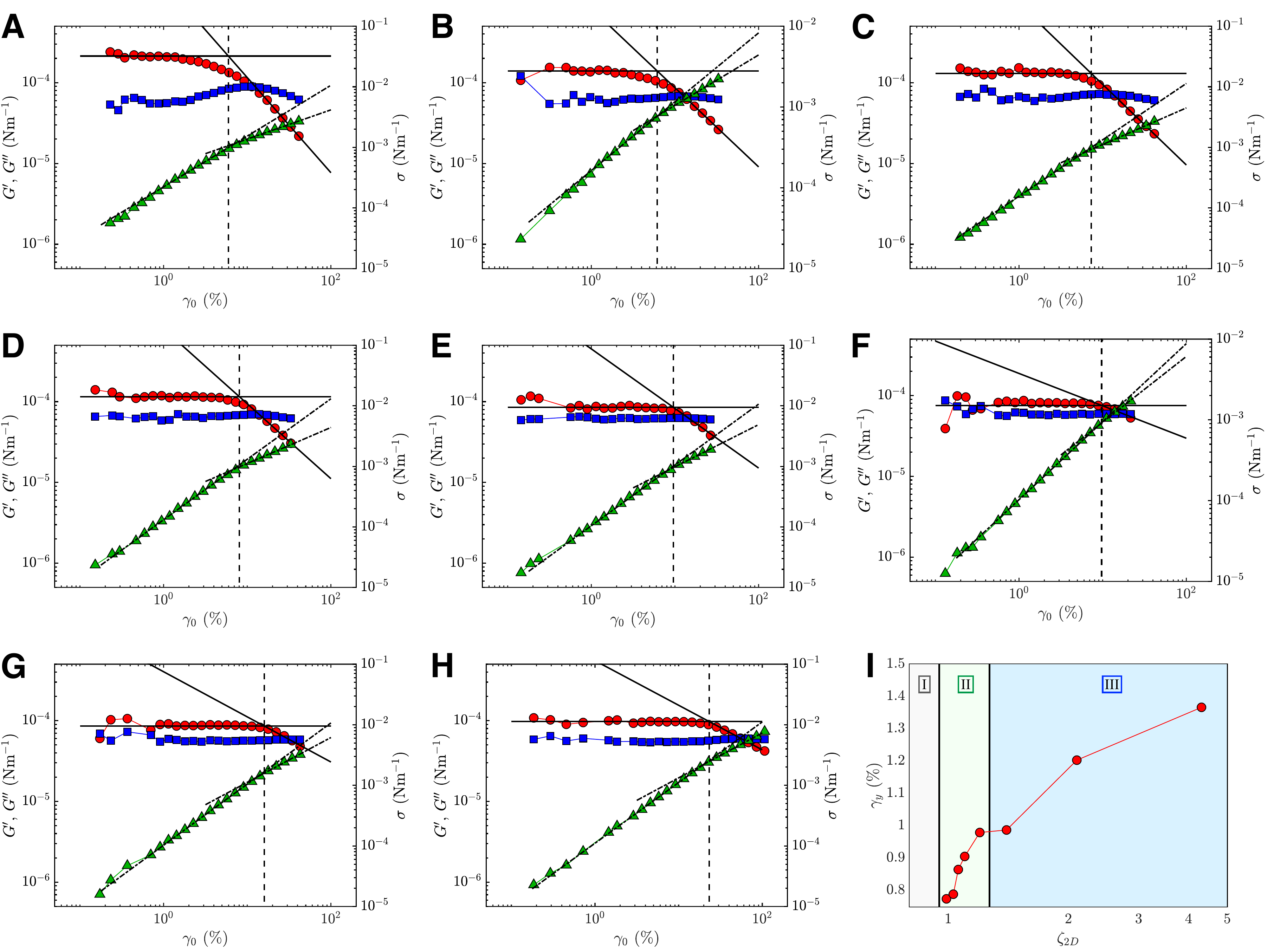}
    \caption{\textbf{Amplitude sweep curves of ULC microgel monolayers at the oil-water interface.}
    (\textbf{A-H}) Values of the elastic (circles) and loss modulus (squares) and stress (triangle, right axes) \emph{vs.}~strain, $\gamma_0$, measured in amplitude sweep ($\omega = 1$~rad s$^{-1}$) for the soft microgels at different values of $\zeta_{2D}$: (\textbf{A}) $0.99\pm0.21$, (\textbf{B}) $1.03\pm0.22$, (\textbf{C}) $1.06\pm0.22$, (\textbf{D}) $1.10\pm0.23$, (\textbf{E}) $1.20\pm0.25$, (\textbf{F}) $1.40\pm0.30$, (\textbf{G}) $2.10\pm0.44$, and (\textbf{H}) $4.31\pm0.92$. Dash-dotted lines are linear fits of $\sigma$~vs~$\gamma_0$ for high and low values of $\gamma_0$. Solid lines are used to evaluate the interception between the different regimes in the course of $G'$. The dashed vertical lines mark the values of $\gamma_y$. (\textbf{I}) Values of $\gamma_y$ as a function of $\zeta_{2D}$.}
    \label{fig:ULC3AS}
\end{figure}

Amplitude sweep measurements at controlled surface pressures were conducted with an angular frequency of 1.0~rad s$^{-1}$. 
The applied shear strain amplitude was varied from 0.5\% to 100\%.
The data in Fig.~\ref{fig:ULC3AS} shows on one hand that the monolayer is a solid ($G^\prime>G^{\prime\prime}$ for $\gamma_0 < 10^1$) and on the other that the monolayer is in the linear viscoelastic regime for $\gamma_0 = 1$\% ($\omega = 1$~rad s$^{-1}$).
In contrast to what observed for monolayer of harder microgels \cite{Sch23}, here $G^\prime$ remains constant for low $\gamma_0$ and then decreases without showing a maximum.
$G^{\prime\prime}$ remains almost constant withing the measured range of $\gamma_0$.
The triangles shows the course of $\sigma$~vs.~$\gamma_0$.
Two different slopes of the data can be identified and fitted with two different power laws (dash-dotted lines).
Conventionally, the value of $\gamma_0$, at which low- and high strain power law variations cross, is considered to be the threshold strain amplitude between liquid-like and solid-like behaviour, $\gamma_y$.

However, as can be seen in some of the curves, e.g.~\ref{fig:ULC3FC_SI}(e)-to-(h), the two different linear regimes are not clearly distinguishable.
In these cases, we obtain the values of $\gamma_y$ considering the course of $G'$: the intercept between the flat part of $G'$ at low $\gamma_0$ and the slope describing the course of $G'$ at high $\gamma_0$ (solid lines) is considered as the critical value of $\gamma_0$ above which the monolayer flows.
The values of $\gamma_y$ determined in this way are marked by the dashed vertical line in panels~\ref{fig:ULC3FC_SI}(a)-to-(h) and their values are shown as a function of $\zeta_{2D}$ (circles) in panel \ref{fig:ULC3FC_SI}(i).
The values determined with this methods, dashed vertical lines in \ref{fig:ULC3FC_SI}(a)-to-(h), coincide with the values where the intercept between the two linear regimes in the curves $\sigma$~vs.~$\gamma_0$ intersect (dash-dotted lines).

\section{Oscillatory frequency sweeps}

\begin{figure}[htbp!]
    \centering
    \includegraphics[width=\textwidth]{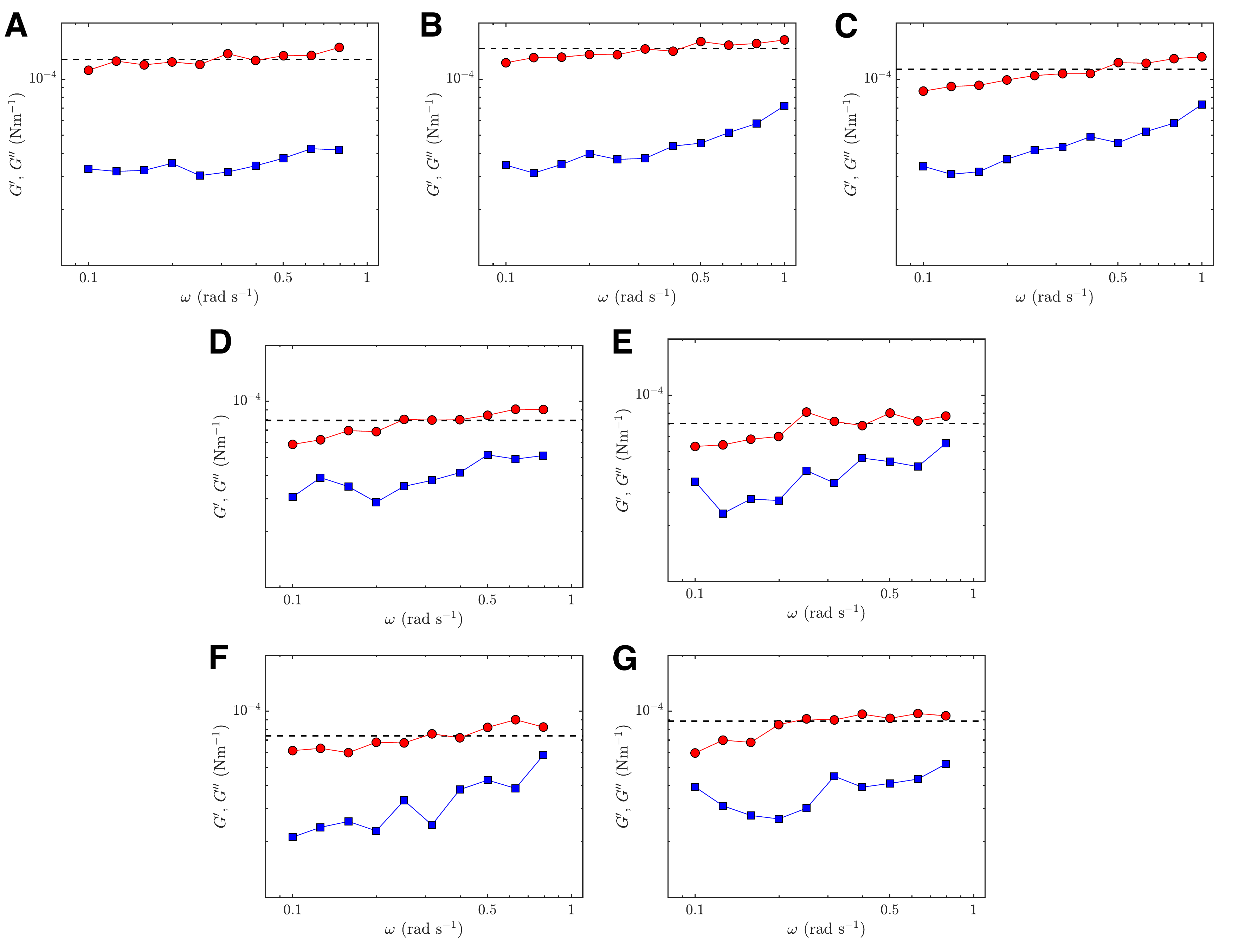}
    \caption{\textbf{Frequency sweep curves of ULC microgel monolayers at the oil-water interface.} (\textbf{A-G})
    Values of the elastic (circles) and loss modulus (squares) \emph{vs.}~the angular frequency, $\omega$, measured in frequency sweep ($\gamma_0 = 1$\%) for the soft microgels at different values of $\zeta_{2D}$: (\textbf{A}) $0.99\pm0.21$, (\textbf{B}) $1.03\pm0.22$, (\textbf{C}) $1.10\pm0.23$, (\textbf{D}) $1.20\pm0.25$, (\textbf{E}) $1.40\pm0.30$, (\textbf{F}) $2.10\pm0.44$, (\textbf{G}) $4.31\pm0.92$. Dashed lines: average values of the elastic modulus.} 
    \label{fig:ULC3_OFS}
\end{figure}

Frequency sweep measurements at controlled surface pressures were performed with a shear strain amplitude of 1\%, i.e.~in the linear viscoelastic region (see Fig.~\ref{fig:ULC3AS}). 
All the measured frequency sweeps in Fig.~\ref{fig:ULC3_OFS} confirms that the monolayer remain solid within the concentrations proved in this study.
We can see that the distance between $G^\prime$ and $G^{\prime\prime}$, first is quite large \ref{fig:ULC3_OFS}(a) and \ref{fig:ULC3_OFS}(b), then decrease, \ref{fig:ULC3_OFS}(c)-to-(e), and finally starts to increase again \ref{fig:ULC3_OFS}(f) and \ref{fig:ULC3_OFS}(g).
In contrast to the case of microgels in bulk \cite{Sco21} and to the monolayer of harder microgels \cite{Sch23} we cannot find a clear minimum in $G^{\prime\prime}$ at a well defined frequency $\omega_m$.
This makes impossible to compute the value of the plateau of $G^\prime$ as the value corresponding to $\omega_m$: $G_p = G^\prime(\omega_m)$.
Therefore here, we choose to use the average of the value of $G^\prime$ as value of $G_p$ and use the standard deviation as error.
The values of $G_p$ as a function of $\zeta_{2D}$ are plotted as circles in Fig.~4.

\newpage

\section{Oscillatory shear of multi-Hertzian Discs}

\begin{figure}[htbp!]
    \centering
    \includegraphics[width=0.8\textwidth]{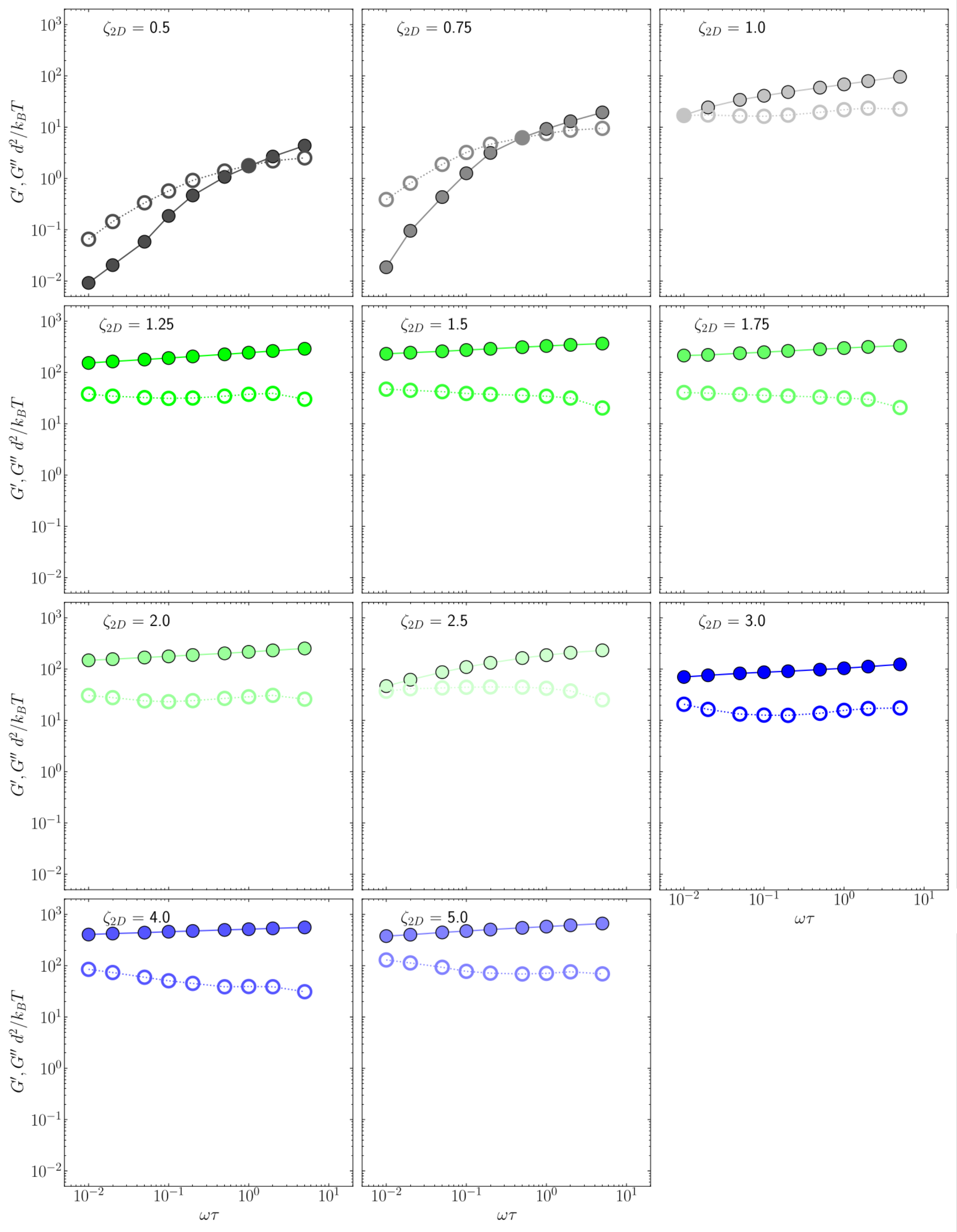}
    \caption{\textbf{Simulated frequency sweeps of multi-Hertzian discs.} Simulated storage (solid symbols) and loss modulus (hollow symbols) with a strain amplitude of 3~\% as a function of angular frequency for area fractions between 0.5 and 5.0. Colors refer to amplitude sweeps in the different regimes.}
    \label{fig:SIM_G_FS_ind}
\end{figure}

We  apply an oscillatory shear rate to a monolayer of discs interaction with a multi-Hertzian potential, where the shear has the function of a sine wave, with an amplitude of $\gamma_{0} = 0.03$, to stay in the linear visco-elastic regime.
The angular frequency, $\omega$, is then varied from $1\cdot10^{-2}\tau^{-1}$ to $5\cdot10^{0}\tau^{-1}$.
Frequency sweeps as a function of area fraction are reported in \ref{fig:SIM_G_FS_ind}.

\newpage

\section{Particle size at the interface}\label{sec:z_2d}

\begin{figure}[htb!]
    \centering
    \includegraphics[width=0.7\textwidth]{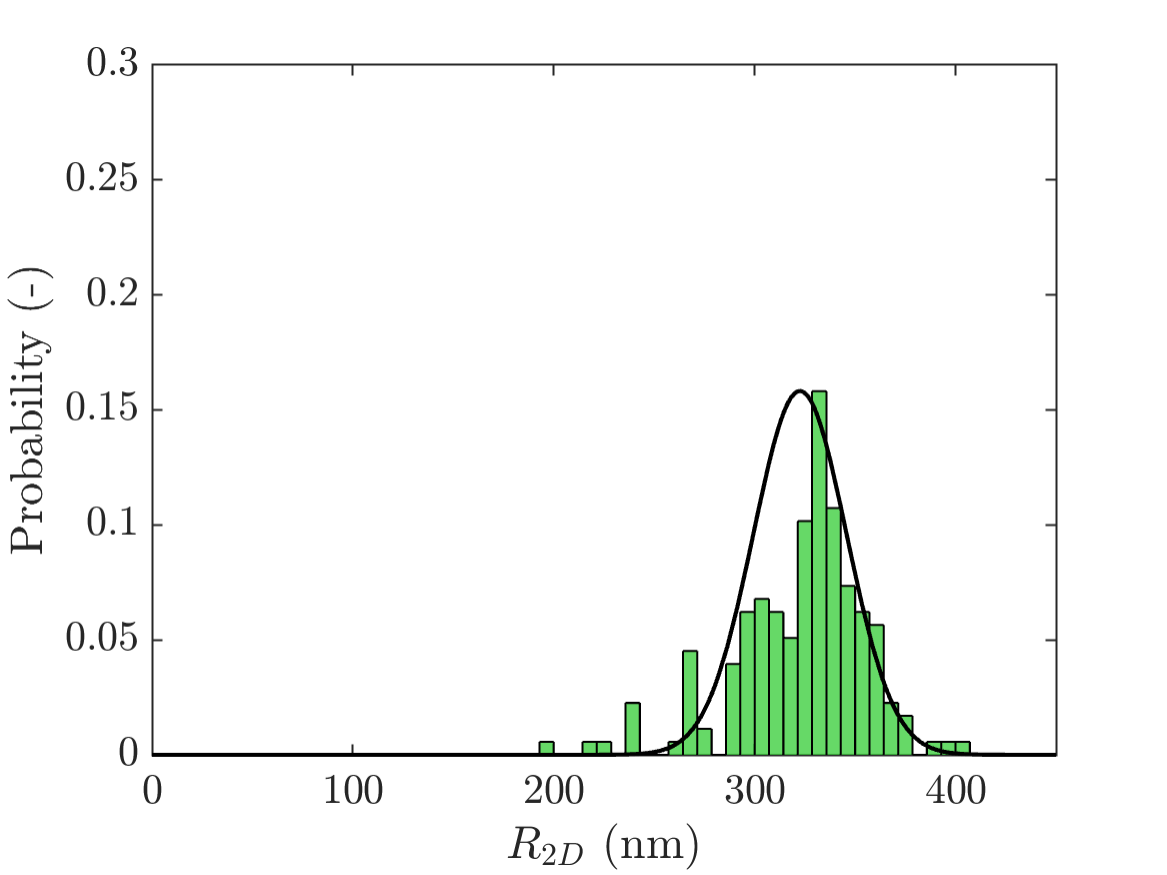}
    \caption{\textbf{Microgel size distribution from AFM}. Distribution of the values of the radius of the used ULC microgels determined from the analysis of AFM images. The solid line represents it fitted by a Gaussian.}
    \label{fig:R_dist}
\end{figure}

The value of the particle radius used in Eq. 1 is the average radius of the adsorbed microgels at the interface before they get in contact to each other.
The distribution of the values of the radii for the ULC microgels, as obtained from the analysis of the AFM micrographs with the microgels in this condition, is shown in Fig.~\ref{fig:R_dist}.
The data are obtained considering $\gtrsim 220$ microgels imaged before contact using AFM.
To obtain the value of the average radius, the data in Fig.~\ref{fig:R_dist} are fitted using a Gaussian, solid line.
From the firs a value for the radius $R_{2D} = 323\pm33$~nm is obtained.

\bibliographystyle{apsrev4-1}

\end{document}